\documentclass[preprint,12pt]{elsarticle}

\usepackage[T1]{fontenc}
\usepackage[utf8]{inputenc}
\usepackage[scaled=0.92]{inconsolata}
\usepackage{textcomp}
\usepackage{amsmath,amssymb,amsfonts,bm}

\usepackage{graphicx}
\usepackage{float}
\usepackage[table]{xcolor}
\usepackage{subfig}
\usepackage{caption}
\usepackage{adjustbox}
\usepackage{multirow}
\usepackage{makecell}
\usepackage{booktabs}
\usepackage{array}

\usepackage{doi}
\usepackage{url}
\usepackage{enumitem}

\usepackage{listings}

\usepackage{tikz}
\usepackage{tikz-feynman}
\tikzfeynmanset{compat=1.1.0,warn luatex=false}
\usetikzlibrary{shapes.geometric,arrows,arrows.meta,fit,positioning,shadows.blur}

\usepackage[edges]{forest}

\usepackage[most]{tcolorbox}

\usepackage{dirtree}
\usepackage{eso-pic}
\definecolor{VLQBlue}{RGB}{24,78,119}
\definecolor{VLQTeal}{RGB}{0,150,136}
\definecolor{VLQMagenta}{RGB}{136,14,79}
\definecolor{VLQAmber}{RGB}{255,143,0}
\definecolor{VLQGray}{RGB}{84,110,122}
\definecolor{VLQInk}{RGB}{7,29,43}
\definecolor{VLQBg}{RGB}{248,249,251}
\definecolor{VLQSoftBlue}{RGB}{238,245,251}
\definecolor{VLQSoftGreen}{RGB}{230,245,235}
\definecolor{VLQRed}{RGB}{180,30,30}
\definecolor{VLQGold}{RGB}{180,140,0}

\usepackage{hyperref}
\hypersetup{
	colorlinks=true,
	linkcolor=VLQBlue!70!black,
	citecolor=VLQMagenta!65!black,
	urlcolor=VLQTeal!60!black
}
\usepackage{cleveref}
\usepackage{needspace}
\usepackage{placeins}
\usepackage{etoolbox}

\journal{Computer Physics Communications}
\biboptions{sort&compress}

\BeforeBeginEnvironment{lstlisting}{\Needspace{20\baselineskip}}

\newcommand{\vlqb}{\texttt{VLQBounds}}
\newcommand{\code}[1]{\texttt{#1}}
\newcommand{\githabrepo}{\url{https://github.com/MDBKD/VLQBounds}}

\newcommand{\safeincludegraphics}[2][]{%
	\IfFileExists{#2}{%
		\includegraphics[#1]{#2}%
	}{%
		\fbox{\parbox[c][4.2cm][c]{0.86\linewidth}{\centering\footnotesize Missing figure file: \texttt{\detokenize{#2}}}}%
	}%
}

\lstdefinestyle{vlq-light}{
	backgroundcolor=\color{VLQBg},
	basicstyle=\linespread{1.07}\ttfamily\footnotesize,
	commentstyle=\color{VLQTeal!85!black}\itshape,
	keywordstyle=\bfseries\color{VLQBlue!90!black},
	stringstyle=\color{VLQAmber!80!black},
	numberstyle=\scriptsize\color{VLQGray!80!white},
	numbers=left,
	numbersep=10pt,
	showstringspaces=false,
	frame=single,
	rulecolor=\color{VLQGray!40},
	frameround=tttt,
	breaklines=true,
	tabsize=2,
	upquote=true,
	moredelim=[is][\bfseries\color{VLQMagenta!80!black}]{§}{§},
	emphstyle=\bfseries\color{VLQMagenta!80!black},
	captionpos=b
}

\lstdefinestyle{vlq-terminal}{
	backgroundcolor=\color{VLQInk},
	basicstyle=\linespread{1.07}\ttfamily\footnotesize\color{white},
	commentstyle=\color{VLQTeal!80!white}\itshape,
	keywordstyle=\bfseries\color{VLQAmber},
	stringstyle=\color{VLQGold},
	numbers=none,
	showstringspaces=false,
	frame=single,
	rulecolor=\color{VLQGray!40},
	frameround=tttt,
	breaklines=true,
	tabsize=2,
	upquote=true,
	captionpos=b
}

\forestset{
	repository tree/.style={
		for tree={
			grow'=east,
			parent anchor=east,
			child anchor=west,
			anchor=west,
			edge={-Latex, semithick, draw=VLQGray!65},
			l sep=14pt,
			s sep=6pt,
			inner sep=2pt,
			align=left,
			font=\ttfamily\scriptsize
		}
	},
	repo folder/.style={
		draw=VLQBlue!65!black,
		fill=VLQBlue!7,
		rounded corners=2pt,
		text=VLQBlue!80!black,
		font=\ttfamily\bfseries\scriptsize,
		inner xsep=4pt,
		inner ysep=2.5pt
	},
	repo module/.style={
		draw=VLQTeal!70!black,
		fill=VLQTeal!8,
		rounded corners=2pt,
		text=VLQTeal!55!black,
		font=\ttfamily\bfseries\scriptsize,
		inner xsep=4pt,
		inner ysep=2.5pt
	},
	repo file/.style={
		draw=VLQGray!60!black,
		fill=VLQGray!8,
		rounded corners=2pt,
		text=black,
		font=\ttfamily\scriptsize,
		inner xsep=3.5pt,
		inner ysep=2pt
	}
}

\begin{document}
\AddToShipoutPictureFG*{%
	\AtPageUpperLeft{%
		\put(\LenToUnit{\paperwidth-4.4cm},\LenToUnit{-1.35cm}){%
			\parbox[t]{3.8cm}{\raggedleft\normalfont\small IFJPAN-IV-2026-9}%
		}%
	}%
}
\begin{frontmatter}

\title{\vlqb: Confronting Vector-Like Quark Models with LHC Searches}

\author[tangier,france]{A. Arhrib}
\ead{ aarhrib@gmail.com}
\author[safi]{R. Benbrik}
\ead{r.benbrik@uca.ac.ma}
\author[ifj]{M. Boukidi}
\ead{mohammed.boukidi@ifj.edu.pl}
\author[safi]{M. Ech-chaouy}
\ead{m.echchaouy.ced@uca.ac.ma}
\author[uu,soton]{S. Moretti}
\ead{stefano.moretti@physics.uu.se}
\ead{s.moretti@soton.ac.uk}
\author[essa]{K. Kahime}
\ead{Kahimek@gmail.com}
\author[safi]{K. Salime}
\ead{k.salime.ced@uca.ac.ma}
\author[cas,ucas]{Q.S. Yan}
\ead{yanqishu@ucas.ac.cn}
\address[tangier]{Abdelmalek Essaadi University, FST Tanger B.P. 416, Morocco}
\address[france]{LAPTh, CNRS, Université Savoie Mont-Blanc, 9 Chemin de Bellevue, 74940, Annecy, France}
\address[safi]{Polydisciplinary Faculty, Laboratory of Physics, Energy, Environment, and Applications, Cadi Ayyad University, Safi, Morocco}
\address[ifj]{Institute of Nuclear Physics, Polish Academy of Sciences, ul. Radzikowskiego 152, Cracow 31-342, Poland}
\address[uu]{Department of Physics \& Astronomy, Uppsala University, SE-751 20 Uppsala, Sweden}
\address[soton]{School of Physics \& Astronomy, University of Southampton, Southampton SO17 1BJ, United Kingdom}
\address[essa]{Laboratoire Interdisciplinaire de Recherche en Environnement, Management, Energie et Tourisme (LIREMET), ESTE, Cadi Ayyad University, B.P. 383, Essaouira, Morocco}
\address[cas]{Center for Future High Energy Physics, Chinese Academy of Sciences, Beijing 100049, China}
\address[ucas]{School of Physics Sciences, University of Chinese Academy of Sciences, Beijing 100039, China}

\begin{abstract}
\noindent
We present \vlqb, a public, data-driven Python framework for testing Vector-Like Quark (VLQ) scenarios against Large Hadron Collider (LHC) exclusion limits from ATLAS and CMS. The framework incorporates public results on both pair and single VLQ production and supports the main parameterisations used in experimental interpretations, including mass-mixing, mass-coupling, and mass-width representations. For each parameter point, the predicted cross-section or effective coupling is compared channel by channel to the corresponding observed and expected experimental limits through interpolation over machine-readable grids. The most sensitive analysis is automatically identified and a 95\% Confidence-Level exclusion verdict is returned, together with the observed and expected sensitivity ratios and the metadata needed for reproducible reinterpretation. The modular structure of \vlqb\ makes it suitable for fast phenomenological scans, validation of public limits, and future extensions to new collider searches and non-minimal VLQ decay patterns.
\end{abstract}


\end{frontmatter}

	\section{Introduction}
	\label{sec:intro}
	
	The exploration of physics Beyond the Standard Model (BSM) remains a central focus of the high-energy physics programme at the LHC. Despite its remarkable successes, the SM does not address several open questions, including the origin of the fermion mass hierarchy, the detailed structure of Electro-Weak Symmetry Breaking (EWSB), the presence of Dark Matter (DM), and the mechanism behind the observed matter-antimatter asymmetry in the Universe. Among the most actively studied classes of extensions are models that introduce VLQs, hypothetical spin-$1/2$ fermions whose left- and right-handed components transform identically under the SM gauge group~\cite{Aguilar-Saavedra:2013qpa,Buchkremer:2013bha,Fuks:2016ftf,Alves:2023ufm}.
	
	Such VLQs arise naturally in a variety of ultraviolet completions, including Composite Higgs Models (CHMs)~\cite{Agashe:2004rs,Contino:2006qr}, Little Higgs constructions~\cite{Arkani-Hamed:2002ikv,Schmaltz:2005ky}, extra-dimensional scenarios~\cite{Randall:1999ee,Carena:2007tn}, and Grand Unified Theories (GUTs) with extended gauge symmetries~\cite{Hewett:1988xc}. Beyond these frameworks, the phenomenology of VLQs has also been extensively investigated in the context of the Two-Higgs-Doublet Model (2HDM)~\cite{Arhrib:2024dou,Arhrib:2024tzm,Arhrib:2024mbq,Benbrik:2022kpo,Abouabid:2023mbu,Benbrik:2024bxt,Benbrik:2024hsf,Arhrib:2024nbj,Benbrik:2023xlo,Dermisek:2019vkc,Benbrik:2025nfw}. Their vector-like character of these new fermions permits gauge-invariant mass terms independently of EWSB and  without necessarily destabilising precision EW constraints. Depending on their mixing with SM quarks, VLQs can modify Higgs and EW couplings, induce flavour-non-universal signatures, and motivate a broad experimental search programme.
	
	At the LHC, VLQs are produced mainly through two complementary mechanisms. Pair production proceeds through QCD interactions and is largely model-independent, depending primarily on the VLQ mass. Single production proceeds through EW  interactions and is directly sensitive to the mixing or effective coupling between the heavy fermion and SM quarks. ATLAS and CMS have searched for the four canonical VLQ states, namely the up-type partner $T$, the down-type partner $B$, and the exotic partners $X$ and $Y$, across a wide range of final states. The resulting constraints now reach the TeV scale in several channels~\cite{Benbrik:2024fku,Banerjee:2024zvg}.
	
	The reinterpretation of these results is not straightforward. Experimental limits are published in several native forms: upper bounds on cross-sections, exclusion contours in the $(m_Q,\sin\theta_{L/R})$ plane, limits in the $(m_Q,\kappa)$ one, and bounds in the $(m_Q,\Gamma_Q/m_Q)$ one under fixed Branching Ratio (BR) assumptions. The diversity of representations, topologies, and assumptions makes it difficult to test arbitrary VLQ scenarios in a coherent and automated way, especially when scanning large parameter spaces or studying non-minimal representations.

	Public tools directly tailored to VLQ interpretations are still relatively scarce. The closest dedicated public code is \texttt{XQCAT}, which determines exclusion confidence levels for scenarios with one or several extra quarks by relying on precomputed efficiencies for QCD pair production and on-shell decays~\cite{Barducci:2014ila}. More general public reinterpretation frameworks, such as \texttt{MadAnalysis~5}, \texttt{CheckMATE}, \texttt{SModelS}, \texttt{Fastlim}, and \texttt{Contur}, can also be used to confront BSM scenarios with collider data~\cite{Conte:2014zja,Dercks:2016npn,Ambrogi:2017neo,Papucci:2014rja,Buckley:2020wzk}. However, these tools are not organised around the native $(m_Q,s_{L/R})$, $(m_Q,\kappa)$, and $(m_Q,\Gamma_Q/m_Q)$ representations in which VLQ limits are often presented. In this sense, \vlqb\ is designed as a dedicated and lightweight layer for the direct use of public VLQ searches, while remaining complementary to broader recasting infrastructures.
	
	To address this issue, we introduce \vlqb, a Python framework that compares theoretical predictions for VLQ models with public ATLAS and CMS exclusion limits. The framework handles the standard VLQ species $T$, $B$, $X$, and $Y$, including singlet, doublet, triplet, and fixed BR  configurations. It accepts input in terms of masses, mixing angles, effective couplings, and/or relative widths and performs the required translations internally. For each model point, \vlqb\ evaluates the relevant observables, interpolates the experimental limits in their native variables, identifies the most sensitive analysis, and returns a 95\% CL exclusion decision together with the corresponding diagnostic information.
	
	The code is intended for parameter scans and validation studies. It includes an interface for reproducing published mass-width and mass-coupling exclusion plots directly from the internal \vlqb\ data files, without hard-coding numerical inputs in external scripts. The same internal workflow is used for validation plots and scan outputs.
	
	The paper is organised as follows. Section~\ref{sec:framework} summarises the theoretical setup and conventions. Section~\ref{sec:approach} describes the exclusion methodology. Section~\ref{sec:limits} summarises the experimental inputs. Section~\ref{sec:architecture} details the software architecture, public API, scan interface, and validation plotting tools. Section~\ref{sec:validation} presents validation examples while Sec.~\ref{sec:results} gives representative applications. Conclusions are presented in Sec.~\ref{sec:conclusion}.
	
	\section{Model Framework and Notation}
	\label{sec:framework}
	
	Here, we summarise only the conventions required by \vlqb\  (a complete derivation of the mass matrices, mixing structure, and gauge/Higgs interactions can be found in Ref.~\cite{Benbrik:2024fku}. With an ${\rm SU}(2)_L$ doublet scalar sector and renormalisable quark-VLQ mixings, the admissible gauge-covariant multiplets are~\cite{delAguila:2000aa}
	\begin{align}
		& T_{L,R}^0\,, \quad B_{L,R}^0 && \text{(singlets)}, \nonumber\\
		& (X\,T^0)_{L,R}\,,\quad (T^0\,B^0)_{L,R}\,,\quad (B^0\,Y)_{L,R} && \text{(doublets)}, \nonumber\\
		& (X\,T^0\,B^0)_{L,R}\,,\quad (T^0\,B^0\,Y)_{L,R} && \text{(triplets)}.
	\end{align}
	The states $Q=\{T,B\}$ carry electric charges $\{+2/3,-1/3\}$, while the exotic partners $X$ and $Y$ carry $+5/3$ and $-4/3$, respectively. We restrict the treatment to one additional VLQ multiplet mixing dominantly with the third SM generation.
	
	\paragraph{Mixing Conventions.}
	In the up- and down-type sectors, weak and mass eigenstates are related by $2\times2$ unitary rotations $U^q_{L,R}$ with angles $\theta^q_{L,R}$. We use $s^q_{L,R}\equiv\sin\theta^q_{L,R}$ and $c^q_{L,R}\equiv\cos\theta^q_{L,R}$. For singlets and triplets the dominant mixing is left-handed, while for doublets it is right-handed. The subdominant angle is fixed by~\cite{Aguilar-Saavedra:2013qpa}
	\begin{align}
		\tan\theta_R^q = \frac{m_q}{m_Q}\tan\theta_L^q \quad &\text{(singlets, triplets)}, 
		\qquad
		\tan\theta_L^q = \frac{m_q}{m_Q}\tan\theta_R^q \quad \text{(doublets)},
		\label{eq:angle-relations}
	\end{align}
	with $(q,m_q,m_Q)=(u,m_t,m_T)$ or $(d,m_b,m_B)$. Complex phases are neglected, since they are not relevant for the exclusion logic implemented here.
	
	\paragraph{Decays and Working Assumptions.}
	In the minimal SM$+$VLQ limit, where no additional light BSM states enter the decay chain, the partial widths are determined by the gauge and Higgs interactions fixed by the mixing angles. The standard decay patterns are
	\[
	T\to W^\pm b,\;Zt,\;ht,\qquad
	B\to W^\pm t,\;Zb,\;hb,\qquad
	X\to W^\pm t,\qquad
	Y\to W^\pm b.
	\]
	When an experimental analysis is published under fixed BR assumptions, \vlqb\ uses those assumptions only for the corresponding comparison.
	
	\paragraph{Input Parameterisations.}
	For each VLQ species, the framework accepts one of the following inputs:
	\[
	(m_Q,s_{L/R}),\qquad (m_Q,\kappa),\qquad (m_Q,\Gamma_Q/m_Q).
	\]
	Theory predictions for production and decays are evaluated accordingly. When necessary, the code translates between these representations using the relations in Table~\ref{tab:kappa}.
	
	\begin{table}[t]
		\centering
		\renewcommand{\arraystretch}{1.15}
		\begin{tabular}{c|c|c}
			\hline\hline
			\textbf{VLQ representation} & \textbf{Model} & \textbf{Translation: $\kappa$ in terms of mixing} \\
			\hline\hline
			\multirow{3}{*}{$T$} & Singlet & $\sqrt{2}\,s_L$ \\
			& Doublet $(X,T)$ & $s_R\,c_R$ \\
			& Doublet $(T,B)$ & $s_R^{u}\,c_R^{u}$ \\
			\hline
			\multirow{3}{*}{$B$} & Singlet & $\sqrt{2}\,s_L$ \\
			& Doublet $(B,Y)$ & $s_R\,c_R$ \\
			& Doublet $(T,B)$ & $s_R^{d}\,c_R^{d}$ \\
			\hline
			$X$ & Doublet $(X,T)$ & $s_R$ \\
			\hline
			$Y$ & Doublet $(B,Y)$ & $s_R$ \\
			\hline
			$Y$ & Triplet $(T,B,Y)$ & $\sqrt{2}\,s_L^{d}$ \\
			\hline\hline
		\end{tabular}
		\caption{Map between the experimental coupling parameter $\kappa$ and the dominant quark-VLQ mixing angles used in this work.}
		\label{tab:kappa}
	\end{table}
	
	\paragraph{Masses and Multiplet Consistency.}
	By default, \vlqb\ treats the masses of the different VLQ species as independent user-specified inputs. The conventions above are sufficient to map arbitrary points in VLQ parameter space onto ATLAS and CMS limit grids in their native variables, whether cross-section, $(m_Q,s_{L/R})$, $(m_Q,\kappa)$, or $(m_Q,\Gamma_Q/m_Q)$. Full details of the interaction Lagrangian, mixing structure, and diagonalisation conventions are given in Ref.~\cite{Benbrik:2024fku}.
	
	\section{\vlqb\ Approach}
	\label{sec:approach}
	
	The absence of experimental evidence for VLQs at the LHC is translated into constraints on their production and decay rates. These constraints are usually expressed as upper limits on cross-sections for processes in which VLQs are produced and subsequently decay into experimentally accessible final states. Two production mechanisms are central: pair production and single production.
	
	Pair production, $pp\to Q\bar Q$, is driven by QCD interactions and is therefore largely independent of the detailed VLQ mixing pattern. The theoretical prediction depends predominantly on the VLQ mass, making pair production the natural baseline for model-independent exclusions. Experimental upper bounds on $\sigma(pp\to Q\bar Q)$ can therefore be recast into lower limits on $m_Q$ under specified BR hypotheses.
	
\begin{figure}[t]
		\centering
		\subfloat[]{%
			\begin{tikzpicture}[scale=0.92]
				\begin{feynman}
					\vertex (a);
					\vertex [left=1.9cm of a] (b) {$q$};
					\vertex [right=1.9cm of a] (c) {$q'/q$};
					\vertex [below=2.0cm of a] (d);
					\vertex [right=1.9cm of d] (e);
					\vertex [below=2.0cm of d] (f);
					\vertex [left=1.9cm of f] (g) {$g$};
					\vertex [right=1.9cm of f] (k) {$\bar b/\bar t$};
					\diagram*{
						(b) -- [fermion] (a) -- [fermion] (c),
						(a) -- [boson, edge label'=$W/Z$] (d) -- [fermion, line width=0.55mm, edge label=$Q$] (e),
						(f) -- [fermion, edge label'=$b/t$] (d),
						(f) -- [gluon] (g),
						(k) -- [fermion] (f),
					};
				\end{feynman}
		\end{tikzpicture}}
		\hspace{0.8cm}
		\subfloat[]{%
			\begin{tikzpicture}[scale=0.92]
				\begin{feynman}
					\vertex (a);
					\vertex [left=1.9cm of a] (b) {$q$};
					\vertex [right=1.9cm of a] (c) {$q'/q$};
					\vertex [below=1.5cm of a] (d);
					\vertex [below=1.5cm of d] (e);
					\vertex [right=1.9cm of e] (x) {$H/Z$};
					\vertex [below=1.5cm of e] (f);
					\vertex [right=1.9cm of d] (y) {$b/t$};
					\vertex [left=1.9cm of f] (g) {$g$};
					\vertex [right=1.9cm of f] (k) {$\bar b/\bar t$};
					\diagram*{
						(b) -- [fermion] (a) -- [fermion] (c),
						(a) -- [boson, edge label'=$W/Z$] (d),
						(e) -- [fermion, line width=0.55mm, edge label=$Q$] (d),
						(f) -- [fermion, edge label'=$b/t$] (e),
						(e) -- [boson] (x),
						(d) -- [fermion] (y),
						(f) -- [gluon] (g),
						(k) -- [fermion] (f),
					};
				\end{feynman}
		\end{tikzpicture}}
		
		\vspace{0.25cm}
		
		\subfloat[]{%
			\begin{tikzpicture}[scale=0.92]
				\begin{feynman}
					\vertex (a);
					\vertex [above left=1.9cm of a] (b) {$g$};
					\vertex [below left=1.9cm of a] (c) {$g$};
					\vertex [right=1.9cm of a] (d);
					\vertex [above right=1.9cm of d] (e) {$Q$};
					\vertex [below right=1.9cm of d] (f) {$\bar Q$};
					\diagram*{
						(b) -- [gluon] (a) -- [gluon] (c),
						(a) -- [gluon, edge label'=$g$] (d),
						(f) -- [fermion, line width=0.55mm] (d) -- [fermion, line width=0.55mm] (e),
					};
				\end{feynman}
		\end{tikzpicture}}
		\hspace{0.8cm}
		\subfloat[]{%
			\begin{tikzpicture}[scale=0.92]
				\begin{feynman}
					\vertex (a);
					\vertex [above left=1.9cm of a] (b) {$q$};
					\vertex [below left=1.9cm of a] (c) {$\bar q'$};
					\vertex [right=1.9cm of a] (d);
					\vertex [above right=1.9cm of d] (e) {$Q$};
					\vertex [below right=1.9cm of d] (f) {$\bar Q$};
					\diagram*{
						(b) -- [fermion] (a) -- [fermion] (c),
						(a) -- [gluon, edge label'=$g$] (d),
						(f) -- [fermion, line width=0.55mm] (d) -- [fermion, line width=0.55mm] (e),
					};
				\end{feynman}
		\end{tikzpicture}}
		\caption{Representative VLQ production topologies: single production in panels (a) and (b), and pair production in panels (c) and (d).}
		\label{fig:prod_diagrams}
	\end{figure}
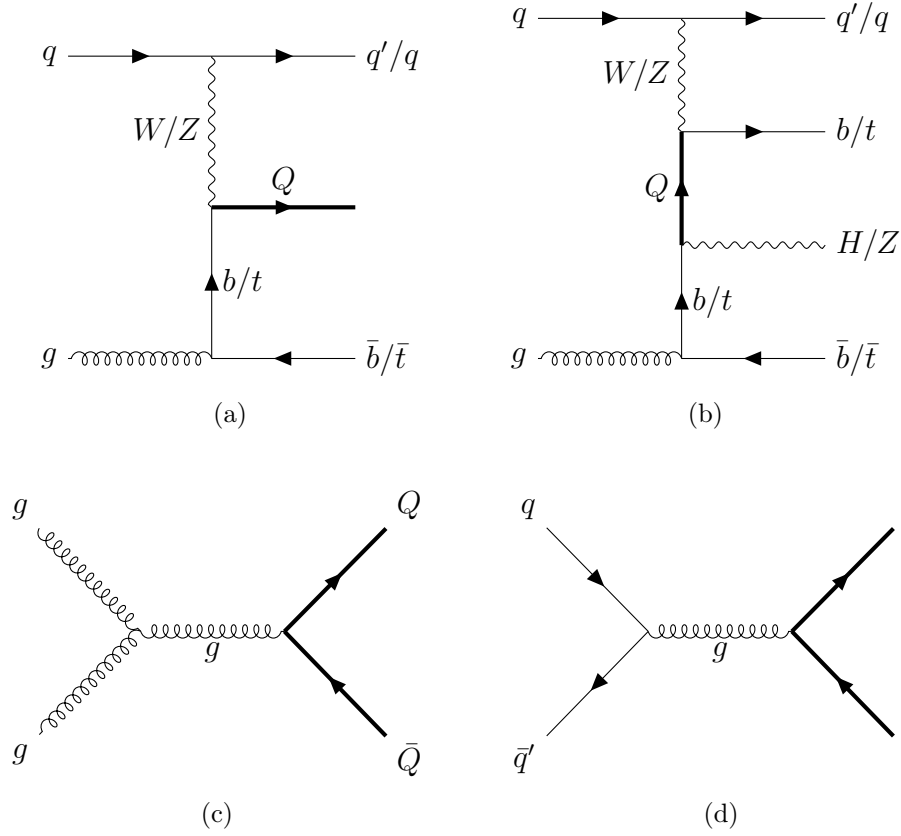

	Single production, by contrast, is mediated by EW interactions and depends on both the VLQ mass and  effective couplings to SM quarks. The corresponding limits are therefore interpreted in a two-dimensional space, typically $(m_Q,\kappa)$ or $(m_Q,\Gamma_Q/m_Q)$. Representative single- and pair-production topologies are shown in Fig.~\ref{fig:prod_diagrams}.
	
	The \vlqb\ approach combines
	\begin{enumerate}[leftmargin=1.25cm]
		\item experimental upper limits on final-state signatures,
		\item theoretical predictions for pair- and single-production rates, and
		\item BRs into the experimentally relevant decay modes.
	\end{enumerate}
	For a given model point, the code maps the user input onto the appropriate experimental variables, evaluates the relevant theory prediction, and compares it with the published observed and expected limits. This allows exclusion regions to be derived without a detector-level reanalysis.
	
	LHC searches for VLQs cover a broad range of theoretical scenarios, from simplified models in which a single VLQ decays into a fixed SM final state to complete EW multiplets such as singlets, doublets, or triplets. Current analyses exclude VLQ masses up to the TeV scale, depending on the assumed decay pattern and production mechanism. The limits arise from searches involving $W$, $Z$, or Higgs bosons in association with heavy-flavour quarks as well as multilepton and boosted-object analyses designed to enhance sensitivity to heavy resonances.
	
	To facilitate reinterpretation, the relevant searches are implemented in the \vlqb\ framework. The included analyses are summarised in Tables~\ref{tab:1}-\ref{tab:6}. For each search, the relevant theoretical prediction $\sigma_{\rm theo}$ is compared with the corresponding observed and expected limits, $\sigma_{\rm obs}$ and $\sigma_{\rm exp}$, using the native variables of the experimental publication.
	
	\FloatBarrier
	\section{Experimental Datasets}
	\label{sec:limits}
	
	The implemented dataset collection includes ATLAS and CMS searches relevant to pair and single production of VLQs at the LHC. The searches cover top partners $T$, bottom partners $B$, and the exotic partners $X$ and $Y$, with decay topologies involving $W$, $Z$, and the SM Higgs boson in association with heavy-flavour quarks. The implementation is organised through machine-readable JSON inputs supplemented by metadata specifying the experiment, centre-of-mass energy, integrated luminosity, process, decay topology, lepton multiplicity, and native limit representation.
	
	Because published limits are provided at discrete mass, coupling, and/or width values, interpolation is required to evaluate arbitrary model points. One-dimensional curves are treated by shape-preserving or linear interpolation in the published mass variable. Two-dimensional surfaces are interpolated in the native plane, such as $(m_Q,\kappa)$ or $(m_Q,\Gamma_Q/m_Q)$. The framework does not extrapolate beyond the published kinematic coverage, rather, points outside the available domain are flagged.
	
	\subsection{Methodology}
	\label{sec:method}
	
	\paragraph{Signal Hypotheses.}
	We consider both QCD pair production,
	\[
	pp\to Q\bar Q,
	\]
	and EW single production,
	\[
	pp\to Qq,
	\]
	where $Q\in\{T,B,X,Y\}$ and $q$ denotes a SM quark. Pair production depends predominantly on $m_Q$, while single production depends on $m_Q$ and a coupling-like parameter expressed as $s_{L/R}$, $\kappa$, or $\Gamma_Q/m_Q$.
	
	\paragraph{Test Statistics and Decision Rule.}
	For each experimental channel $I$ with observed and expected 95\% CL upper limits $\sigma^I_{\rm obs}$ and $\sigma^I_{\rm exp}$, we compute
	\begin{align}
		r^I_{\rm obs}=\frac{\sigma^I_{\rm theo}}{\sigma^I_{\rm obs}},\qquad
		r^I_{\rm exp}=\frac{\sigma^I_{\rm theo}}{\sigma^I_{\rm exp}},
		\label{equ:obs_rat}
	\end{align}
	where $\sigma^I_{\rm theo}$ is the corresponding theory prediction. The most sensitive channel is selected by maximising $r^I_{\rm exp}$, with ties resolved through $r^I_{\rm obs}$. The point is excluded at 95\% CL if the observed ratio of the selected channel satisfies $r_{\rm obs}\geq1$. This avoids combining statistically independent experimental analyses without the full likelihood information.
	
	\begin{table}[H]
		\centering
		\renewcommand{\arraystretch}{0.7}
		\begin{adjustbox}{max width=0.95\textwidth}
			\begin{tabular}{c|c|c|c|c|c}
				\hline\hline
				\textbf{VLQ} & \textbf{ATLAS} & $\sqrt{s}$ & \textbf{Luminosity} & \textbf{Decay} & \textbf{Lepton multiplicity} \\
				\hline\hline
				& \cite{ATLAS:2017vdo} &13 TeV & $36.1$ fb$^{-1}$ & $Z(\nu\nu)t$ & $1\ell$ \\
				&\cite{ATLAS:2017nap} & 13 TeV & $36.1$ fb$^{-1}$ & $W^\pm b$ & $0\ell$ \\
				&\cite{ATLAS:2018alq} & 13 TeV & $36.1$ fb$^{-1}$ & Inclusive & $2\ell$ SS \\
				&\cite{ATLAS:2018dyh} & 13 TeV & $36.1$ fb$^{-1}$ & W$(\ell\nu)b$ & $1\ell$ \\
				&\cite{ATLAS:2022ozf} & 13 TeV & $139$ fb$^{-1}$ & $H(bb)t$ & $0\ell$ \\
				& \cite{ATLAS:2022tla} & 13 TeV & $139$ fb$^{-1}$ & Inclusive & $1\ell$\\
				& \cite{ATLAS:2018cye}& 13 TeV & $36.1$ fb$^{-1}$ & $Ht,Zt$ & $1\ell$ \\
				&\cite{ATLAS:2022hnn} & 13 TeV & $139$ fb$^{-1}$ & $Z(\ell\ell)t$ & $2\ell$ OS, $3\ell$ \\
				&\cite{ATLAS:2023pja} & 13 TeV & $139$ fb$^{-1}$ & $H(b\bar b)t,Z(qq)t$ & $1\ell$\\
				&\cite{ATLAS:2023bfh} & 13 TeV & $139$ fb$^{-1}$ & $Z(\ell\ell)t$ & $2\ell$ OS, $3\ell$ \\
				& \cite{ATLAS:2024xne} & 13 TeV & $140$ fb$^{-1}$ & $Z(\nu\nu)t$ & $0\ell$\\
				$T$ Singlet&\cite{ATLAS:2024gyc} & 13 TeV & $140$ fb$^{-1}$ & Inclusive & $1\ell$ \\
				& \cite{ATLAS:2018cjd} & 13 TeV & $36.1$ fb$^{-1}$ & $Z(\nu\nu)t$ & $0\ell,1\ell$ \\
				&\cite{ATLAS:2016ovj} & 13 TeV & $3.2$ fb$^{-1}$ & $W^\pm b$ & $1\ell$\\
				& \cite{ATLAS:2016btu} & 13 TeV & $13.2$ fb$^{-1}$ & Inclusive & $1\ell,0\ell$ \\
				& \cite{ATLAS:2016sno} & 13 TeV & $3.2$ fb$^{-1}$ & Inclusive & $2\ell$ SS, $3\ell$\\
				& \cite{ATLAS:2016scx} & 8 TeV & $20.3$ fb$^{-1}$ & $W(\ell\nu)b$ & $1\ell$ \\
				&\cite{ATLAS:2014vpn} & 8 TeV & $20.3$ fb$^{-1}$ & $Zt$ & $2\ell$ OS, $3\ell$\\
				& \cite{ATLAS:2015ktd} & 8 TeV & $20.3$ fb$^{-1}$ & Inclusive & $1\ell$ \\
				&\cite{ATLAS:2015uaw} & 8 TeV & $20.3$ fb$^{-1}$ & Inclusive & $2\ell$ SS \\
				&\cite{ATLAS:2018ziw} & 13 TeV & $36.1$ fb$^{-1}$ & Inclusive & $1\ell,2\ell$ SS, $3\ell$ \\
				& \cite{ATLAS:2015vzd} & 8 TeV & $20.3$ fb$^{-1}$ & Inclusive & $1\ell$ \\
				& \cite{ATLAS:2024xdc} & 13 TeV &$139$ fb$^{-1}$ & $Ht,Zt$ & $0\ell,1\ell,\geq2\ell$\\
				& \cite{ATLAS:2018tnt} & 13 TeV & $36.1$ fb$^{-1}$ & $Zt$ & $2\ell$ OS \\
				& \cite{ATLAS:2024kgp} & 13 TeV & $139$ fb$^{-1}$& $W^\pm b$ & $0\ell$\\
				& \cite{ATLAS:2025bzt} & 13 TeV & $140$ fb$^{-1}$& W$(\ell\nu)b$ & $1\ell$\\
				& \cite{ATLAS:2025tja} & 13 TeV & $140$ fb$^{-1}$& W$(\ell\nu)b$ & $1\ell$\\
				\hline
				& \cite{ATLAS:2017vdo} & 13 TeV & $36.1$ fb$^{-1}$ & $Z(\nu\nu)t$ & $1\ell$ \\
				& \cite{ATLAS:2022tla} & 13 TeV & $139$ fb$^{-1}$ & Inclusive & $1\ell$\\
				&\cite{ATLAS:2018cye}& 13 TeV & $36.1$ fb$^{-1}$ & $Ht,Zt$ & $1\ell$ \\
				&\cite{ATLAS:2022hnn} & 13 TeV & $139$ fb$^{-1}$ & $Z(\ell\ell)t$ & $2\ell$ OS, $3\ell$ \\
				&\cite{ATLAS:2023pja} & 13 TeV & $139$ fb$^{-1}$ & $H(b\bar b)t,Z(qq)t$ & $1\ell$\\
				&\cite{ATLAS:2023bfh} &13 TeV & $139$ fb$^{-1}$ & $Z(\ell\ell)t$ & $2\ell$ OS, $3\ell$ \\
				$T$ Doublet&\cite{ATLAS:2016btu} & 13 TeV & $13.2$ fb$^{-1}$ & Inclusive & $1\ell$\\
				&\cite{ATLAS:2018ziw} & 13 TeV & $36.1$ fb$^{-1}$ & Inclusive & $1\ell,2\ell$ SS, $3\ell$ \\
				&\cite{ATLAS:2014vpn} & 8 TeV & $20.3$ fb$^{-1}$ & $Zt$ & $2\ell$ OS, $3\ell$\\
				& \cite{ATLAS:2015ktd} & 8 TeV & $20.3$ fb$^{-1}$ & Inclusive & $1\ell$ \\
				& \cite{ATLAS:2024xdc} & 13 TeV &$139$ fb$^{-1}$ & $Ht,Zt$ & $0\ell,1\ell,\geq2\ell$\\
				\hline
				& \cite{ATLAS:2017vdo} & 13 TeV & $36.1$ fb$^{-1}$ & $Z(\nu\nu)t$ & $1\ell$ \\
				&\cite{ATLAS:2017nap} & 13 TeV & $36.1$ fb$^{-1}$ & $W^\pm b$ & $0\ell$ \\
				& \cite{ATLAS:2018uky} & 13 TeV & $36.1$ fb$^{-1}$ & Inclusive & $0\ell$\\
				& \cite{ATLAS:2022tla} & 13 TeV & $139$ fb$^{-1}$ & Inclusive & $1\ell$\\
				&\cite{ATLAS:2022hnn} & 13 TeV & $139$ fb$^{-1}$ & $Z(\ell\ell)t$ & $2\ell$ OS, $3\ell$ \\
				BR=100\% & \cite{ATLAS:2016seq} & 13 TeV & $3.2$ fb$^{-1}$ & Inclusive & $1\ell$ \\
				&\cite{ATLAS:2016ovj} & 13 TeV & $3.2$ fb$^{-1}$ & $W^\pm(\ell\nu)b$ & $1\ell$\\
				& \cite{ATLAS:2016btu} & 13 TeV & $13.2$ fb$^{-1}$ & Inclusive & $1\ell,0\ell$ \\
				&\cite{ATLAS:2012qe} & 7 TeV & $4.7$ fb$^{-1}$ & $W^\pm b$ & $0\ell$ \\
				& \cite{ATLAS:2015ktd} & 8 TeV & $20.3$ fb$^{-1}$ & Inclusive & $1\ell$ \\
				&\cite{ATLAS:2012tkh} & 7 TeV & $1.04$ fb$^{-1}$ & $W^\pm b$ & $1\ell$ \\
				&\cite{ATLAS:2024gyc} & 13 TeV & $140$ fb$^{-1}$ & Inclusive & $1\ell$ \\
				& \cite{ATLAS:2018tnt} & 13 TeV & $36.1$ fb$^{-1}$ & $Zt$ & $2\ell$ OS \\
				\hline
			\end{tabular}
		\end{adjustbox}
		\caption{Summary of ATLAS searches for the vector-like top partner. Here, $\ell=e,\mu$ and OS(SS) refers to the Opposite(Same) Sign case.}
		\label{tab:1}
	\end{table}
	
	\begin{table}[H]
		\centering
		\renewcommand{\arraystretch}{0.6}
		\begin{adjustbox}{max width=0.95\textwidth}
			\begin{tabular}{c|c|c|c|c|c}
				\hline\hline
				\textbf{VLQ} & \textbf{ATLAS} & $\sqrt{s}$ & \textbf{Luminosity} & \textbf{Decay} & \textbf{Lepton multiplicity} \\
				\hline\hline
				&\cite{ATLAS:2017nap} & 13 TeV & $36.1$ fb$^{-1}$ & $W^\pm t$ & $0\ell$ \\
				& \cite{ATLAS:2018alq} & 13 TeV & $36.1$ fb$^{-1}$ & Inclusive & $2\ell$ SS \\
				& \cite{ATLAS:2022tla} & 13 TeV & $139$ fb$^{-1}$ & Inclusive & $1\ell$\\
				& \cite{ATLAS:2022hnn} & 13 TeV & $139$ fb$^{-1}$ & $Z(\ell\ell)b$ & $2\ell~{\rm OS},3\ell$ \\
				& \cite{ATLAS:2016sno} & 13 TeV & $3.2$ fb$^{-1}$ & Inclusive & $2\ell$ SS,$3\ell$\\
				$B$ Singlet &\cite{ATLAS:2018ziw} & 13 TeV & $36.1$ fb$^{-1}$ & Inclusive & $1\ell,2\ell~{\rm SS},3\ell$ \\
				&\cite{ATLAS:2014vpn} & 8 TeV & $20.3$ fb$^{-1}$ & $Zb$ & $2\ell~{\rm OS},3\ell$ \\
				& \cite{ATLAS:2015ktd} & 8 TeV & $20.3$ fb$^{-1}$ & Inclusive & $1\ell$ \\
				&\cite{ATLAS:2015uaw} & 8 TeV & $20.3$ fb$^{-1}$ & Inclusive & $2\ell~{\rm SS}$ \\
				& \cite{ATLAS:2023ixh} & 13 TeV & $139$ fb$^{-1}$ & $bH(bb)$ & $0\ell$ \\
				& \cite{ATLAS:2018mpo} & 13 TeV & $36.1$ fb$^{-1}$ & $W^\pm t$ & $1\ell$ \\
				& \cite{ATLAS:2015vzd} & 8 TeV & $20.3$ fb$^{-1}$ & Inclusive & $1\ell$\\
				\hline
				& \cite{ATLAS:2018uky} & 13 TeV & $36.1$ fb$^{-1}$ & Inclusive & $0\ell$\\
				&\cite{ATLAS:2018dyh} & 13 TeV & $36.1$ fb$^{-1}$ & $W^\pm b\,\ell\nu$ & $1\ell$ \\
				& \cite{ATLAS:2022tla} & 13 TeV & $139$ fb$^{-1}$ & Inclusive & $1\ell$\\
				$B$ Doublet & \cite{ATLAS:2022hnn} & 13 TeV & $139$ fb$^{-1}$ & $Z(\ell\ell)b$ & $2\ell~{\rm OS},3\ell$ \\
				&\cite{ATLAS:2018ziw} & 13 TeV & $36.1$ fb$^{-1}$ & Inclusive & $1\ell,2\ell~{\rm SS},3\ell$ \\
				&\cite{ATLAS:2014vpn} & 8 TeV & $20.3$ fb$^{-1}$ & $Zb$ & $2\ell~{\rm OS},3\ell$\\
				& \cite{ATLAS:2023ixh} & 13 TeV & $139$ fb$^{-1}$ & $bH(bb)$ & $0\ell$ \\
				\hline
				& \cite{ATLAS:2017nap} & 13 TeV & $36.1$ fb$^{-1}$ & $W^\pm t$ & $0\ell$ \\
				& \cite{ATLAS:2018dyh} & 13 TeV & $36.1$ fb$^{-1}$ & $Z(\ell\ell)b$ & $2\ell~{\rm OS},\geq3\ell$ \\
				& \cite{ATLAS:2018uky} & 13 TeV & $36.1$ fb$^{-1}$ & Inclusive & $0\ell$\\
				BR=100\% & \cite{ATLAS:2022tla} & 13 TeV & $139$ fb$^{-1}$ & Inclusive & $1\ell$\\
				&\cite{ATLAS:2022hnn} & 13 TeV & $139$ fb$^{-1}$ & $Z(\ell\ell)b$ & $2\ell~{\rm OS},3\ell$ \\
				& \cite{ATLAS:2018mpo} & 13 TeV & $36.1$ fb$^{-1}$ & $W^\pm t$ & $1\ell$ \\
				& \cite{ATLAS:2015uhg} & 8 TeV & $20.3$ fb$^{-1}$ & $W^\pm t$ &$1\ell,2\ell$ \\
				& \cite{ATLAS:2015ktd} & 8 TeV & $20.3$ fb$^{-1}$ & Inclusive & $1\ell$ \\
				\hline
			\end{tabular}
		\end{adjustbox}
		\caption{Summary of ATLAS searches for the vector-like bottom partner. Here, $\ell=e,\mu$ and OS(SS) refers to the Opposite(Same) Sign case.}
		\label{tab:2}
	\end{table}
	
	\begin{table}[H]
		\centering
		\renewcommand{\arraystretch}{0.6}
		\begin{adjustbox}{max width=0.95\textwidth}
			\begin{tabular}{c|c|c|c|c|c}
				\hline\hline
				\textbf{VLQ} & \textbf{ATLAS} & $\sqrt{s}$ & \textbf{Luminosity} & \textbf{Decay} & \textbf{Lepton multiplicity} \\
				\hline\hline
				& \cite{ATLAS:2017nap} & 13 TeV & $36.1$ fb$^{-1}$ & $W^\pm t$ & $0\ell$ \\
				&\cite{ATLAS:2018alq} & 13 TeV & $36.1$ fb$^{-1}$ & Inclusive & $2\ell$ SS \\
				$X$ & \cite{ATLAS:2018mpo} & 13 TeV & $36.1$ fb$^{-1}$ & $W^\pm t$ & $1\ell$ \\
				& \cite{ATLAS:2018alq} & 13 TeV & $36.1$ fb$^{-1}$ & $W^\pm t$ & $2\ell$ SS \\
				&\cite{ATLAS:2016sno} & 13 TeV & $3.2$ fb$^{-1}$ & Inclusive & $2\ell$ SS, $3\ell$\\
				&\cite{ATLAS:2015uaw} & 8 TeV & $20.3$ fb$^{-1}$ & Inclusive & $2\ell$ SS \\
				& \cite{ATLAS:2015vzd} & 8 TeV & $20.3$ fb$^{-1}$ & Inclusive & $1\ell$\\
				\hline
				$Y$ & \cite{ATLAS:2017nap} & 13 TeV & $36.1$ fb$^{-1}$ & $W^\pm b$ & $0\ell$ \\
				& \cite{ATLAS:2024gyc}& 13 TeV & $140$ fb$^{-1}$ & Inclusive & $1\ell$ \\
				&\cite{ATLAS:2016scx} & 8 TeV & $20.3$ fb$^{-1}$ & $W^\pm(\ell\nu)b$ & $1\ell$\\
				\hline
				Doublet $X$ & \cite{ATLAS:2022tla} & 13 TeV & $139$ fb$^{-1}$ & $W^\pm t$ & $1\ell$ \\
				\hline
				Doublet $Y$ & \cite{ATLAS:2016scx}& 8 TeV & $20.3$ fb$^{-1}$ & W$(\ell\nu)b$ & $1\ell$ \\
				& \cite{ATLAS:2016ovj} & 13 TeV & $3.2$ fb$^{-1}$ & $W^\pm b$ & $1\ell$ \\
				& \cite{ATLAS:2018dyh} & 13 TeV & $36.1$ fb$^{-1}$ & W$(\ell\nu)b$ & $1\ell$ \\
				&\cite{ATLAS:2016ovj} & 13 TeV & $3.2$ fb$^{-1}$ & $W(\ell\nu)b$ & $1\ell$\\
				& \cite{ATLAS:2024kgp} & 13 TeV & $139$ fb$^{-1}$& $W^\pm b$ & $0\ell$\\
				\hline
				Triplet $Y$ & \cite{ATLAS:2018dyh} & 13 TeV & $36.1$ fb$^{-1}$ & $W^\pm(\ell\nu)b$ & $1\ell$ \\
				&\cite{ATLAS:2025bzt} & 13 TeV & $140$ fb$^{-1}$& W$(\ell\nu)b$ & $1\ell$\\
				\hline
			\end{tabular}
		\end{adjustbox}
		\caption{Summary of ATLAS searches for the exotic VLQs $X$ and $Y$. Here, $\ell=e,\mu$ and OS(SS) refers to the Opposite(Same) Sign case.}
		\label{tab:3}
	\end{table}
	
	\begin{table}[H]
		\centering
		\renewcommand{\arraystretch}{0.8}
		\begin{adjustbox}{max width=0.95\textwidth}
			\begin{tabular}{c|c|c|c|c|c}
				\hline\hline
				\textbf{VLQ} & \textbf{CMS} & $\sqrt{s}$ & \textbf{Luminosity} & \textbf{Decay} & \textbf{Lepton multiplicity} \\
				\hline\hline
				& \cite{CMS:2022yxp} & 13 TeV & $137$ fb$^{-1}$ & $Z(\nu\nu)t$ & $0\ell$ \\
				&\cite{CMS:2016jce} & 13 TeV & $2.3$ fb$^{-1}$ & $H(bb)t$ & $0\ell$ \\
				&\cite{CMS:2016edj} & 13 TeV & $2.3$ fb$^{-1}$& $H(bb)t$ & $0\ell$ \\
				& \cite{CMS:2017ked} & 13 TeV & $2.6$ fb$^{-1}$ & Inclusive & $1\ell$ \\
				& \cite{CMS:2023agg} & 13 TeV & $138$ fb$^{-1}$ & $H(\gamma\gamma)t$ & $0\ell,1\ell$ \\
				& \cite{CMS:2024qdd} & 13 TeV & $138$ fb$^{-1}$ & $Ht,Zt$ & $0\ell$ \\
				& \cite{CMS:2019afi} & 13 TeV & $35.9$ fb$^{-1}$ & $Ht,Zt$ & $0\ell$ \\
				$T$ Singlet & \cite{CMS:2017voh} & 13 TeV & $35.9$ fb$^{-1}$ & $Z(\ell\ell)t$ & $2\ell$ OS \\
				& \cite{CMS:2016ete} & 13 TeV & $2.3$ fb$^{-1}$ & Inclusive & $1\ell$ \\
				& \cite{CMS:2013hwy} & 8 TeV & $19.5$ fb$^{-1}$ & Inclusive & $1\ell$ \\
				& \cite{CMS:2012mir} & 7 TeV & $5.0$ fb$^{-1}$ & $W^\pm(\ell\nu)b$ & $1\ell$\\
				& \cite{CMS:2018zkf} & 13 TeV & $35.9$ fb$^{-1}$ & Inclusive & $1\ell,2\ell~{\rm SS},3\ell$\\
				& \cite{CMS:2022fck} & 13 TeV & $138$ fb$^{-1}$ & Inclusive & $1\ell,2\ell~{\rm SS},3\ell$\\
				& \cite{CMS:2017gsh} & 13 TeV & $2.3$ fb$^{-1}$ &$Zt$ &$0\ell$ \\
				&\cite{CMS:2025zwi} & 13 TeV & $138$ fb$^{-1}$& $Ht$ & $1\ell$\\
				& \cite{CMS:2024bni} & 13 TeV & $138$ fb$^{-1}$ & $W^\pm b,Zt,Ht$ & $1\ell,2\ell~{\rm SS},\geq3\ell$\\
				\hline
				&\cite{CMS:2016jce} & 13 TeV & $2.3$ fb$^{-1}$ & $H(bb)t$ & $0\ell$ \\
				& \cite{CMS:2018zkf} & 13 TeV & $35.9$ fb$^{-1}$ & Inclusive & $1\ell,2\ell~{\rm SS},3\ell$\\
				&\cite{CMS:2016edj} & 13 TeV & $2.3$ fb$^{-1}$& $H(bb)t$ & $0\ell$ \\
				& \cite{CMS:2018wpl} & 13 TeV & $35.9$ fb$^{-1}$ & $Z(\ell\ell)t$ & $2\ell$ OS\\
				$T$ Doublet & \cite{CMS:2017ked} & 13 TeV & $2.6$ fb$^{-1}$ & Inclusive & $1\ell$ \\
				& \cite{CMS:2017gsh} & 13 TeV & $2.3$ fb$^{-1}$ &$Zt$ &$0\ell$ \\
				& \cite{CMS:2017voh} & 13 TeV & $35.9$ fb$^{-1}$ & $Z(\ell\ell)t$ & $2\ell$ OS\\
				& \cite{CMS:2019afi} & 13 TeV & $35.9$ fb$^{-1}$ & $Ht,Zt$ & $0\ell$ \\
				& \cite{CMS:2022fck} & 13 TeV & $138$ fb$^{-1}$ & Inclusive & $1\ell,2\ell~{\rm SS},3\ell$\\
				& \cite{CMS:2024bni} & 13 TeV & $138$ fb$^{-1}$ & $W^\pm b,Zt,Ht$ & $1\ell,2\ell~{\rm SS},\geq3\ell$\\
				\hline
				& \cite{CMS:2017fpk} & 13 TeV & $2.3$ fb$^{-1}$ & $W^\pm(\ell\nu)b$ & $1\ell$ \\
				& \cite{CMS:2018wpl} & 13 TeV & $35.9$ fb$^{-1}$ & $Z(\ell\ell)t$ & $2\ell$ OS\\
				& \cite{CMS:2017ked} & 13 TeV & $2.6$ fb$^{-1}$ & Inclusive & $1\ell$ \\
				& \cite{CMS:2017ynm} & 13 TeV & $35.8$ fb$^{-1}$ & $W^\pm b$ & $1\ell$\\
				& \cite{CMS:2019eqb} & 13 TeV & $36.1$ fb$^{-1}$ & Inclusive & $0\ell$ \\
				BR=100\% & \cite{CMS:2015lzl} & 13 TeV & $19.7$ fb$^{-1}$ & Inclusive & $1\ell,2\ell$ \\
				& \cite{CMS:2012ab} & 7 TeV & $5.0$ fb$^{-1}$ & $W^\pm b$ & $2\ell$ \\
				& \cite{CMS:2015jwh} & 8 TeV & $19.7$ fb$^{-1}$ & $H(bb)t$ & $0\ell$\\
				\hline
			\end{tabular}
		\end{adjustbox}
		\caption{Summary of CMS searches for the vector-like top partner. Here, $\ell=e,\mu$ and OS(SS) refers to the Opposite(Same) Sign case.}
		\label{tab:4}
	\end{table}
	
	\begin{table}[H]
		\centering
		\renewcommand{\arraystretch}{0.8}
		\begin{adjustbox}{max width=0.95\textwidth}
			\begin{tabular}{c|c|c|c|c|c}
				\hline\hline
				\textbf{VLQ} & \textbf{CMS} & $\sqrt{s}$ & \textbf{Luminosity} & \textbf{Decay} & \textbf{Lepton multiplicity} \\
				\hline\hline
				$B$ Singlet & \cite{CMS:2017ked} & 13 TeV & $2.6$ fb$^{-1}$ & Inclusive & $1\ell$ \\
				& \cite{CMS:2018zkf} & 13 TeV & $35.9$ fb$^{-1}$ & Inclusive & $1\ell,2\ell~{\rm SS},3\ell$\\
				& \cite{CMS:2022fck} & 13 TeV & $138$ fb$^{-1}$ & Inclusive & $1\ell,2\ell~{\rm SS},3\ell$\\
				& \cite{CMS:2017gsh} & 13 TeV & $2.3$ fb$^{-1}$ &$Zb$ &$0\ell$ \\
				& \cite{CMS:2018kcw} & 13 TeV & $35.9$ fb$^{-1}$& $H(bb)b$& $0\ell$ \\
				& \cite{CMS:2021mku} & 13 TeV & $138$ fb$^{-1}$ & $W^\pm t$ & $0\ell$ \\
				& \cite{CMS:2024bni} & 13 TeV & $138$ fb$^{-1}$ & $Wt,Zb,Hb$ & $1\ell,2\ell~{\rm SS},\geq3\ell$\\
				& \cite{CMS:2024xbc} & 13 TeV & $138$ fb$^{-1}$ & Inclusive & $0\ell,2\ell$ SS \\
				\hline
				$B$ Doublet & \cite{CMS:2018wpl} & 13 TeV & $35.9$ fb$^{-1}$ & $Z(\ell\ell)b$ & $2\ell$ OS\\
				& \cite{CMS:2017ked} & 13 TeV & $2.6$ fb$^{-1}$ & Inclusive & $1\ell$ \\
				& \cite{CMS:2018zkf} & 13 TeV & $35.9$ fb$^{-1}$ & Inclusive & $1\ell,2\ell~{\rm SS},3\ell$\\
				& \cite{CMS:2022fck} & 13 TeV & $138$ fb$^{-1}$ & Inclusive & $1\ell,2\ell~{\rm SS},3\ell$\\
				& \cite{CMS:2018kcw} & 13 TeV & $35.9$ fb$^{-1}$& $H(bb)b$ & $0\ell$ \\
				& \cite{CMS:2020ttz} & 13 TeV & $138$ fb$^{-1}$& $Hb,Zb$ & $0\ell$ \\
				& \cite{CMS:2024xbc} & 13 TeV & $138$ fb$^{-1}$ & Inclusive & $0\ell,2\ell$ SS \\
				& \cite{CMS:2017gsh} & 13 TeV & $2.3$ fb$^{-1}$ &$Zb$ &$0\ell$ \\
				& \cite{CMS:2024bni} & 13 TeV & $138$ fb$^{-1}$ & $Wt,Zb,Hb$ & $1\ell,2\ell~{\rm SS},\geq3\ell$\\
				\hline
				& \cite{CMS:2018wpl} & 13 TeV & $35.9$ fb$^{-1}$ & $Z(\ell\ell)b$ & $2\ell$ OS\\
				& \cite{CMS:2019eqb} & 13 TeV & $36.1$ fb$^{-1}$ & Inclusive & $0\ell$ \\
				BR=100\% & \cite{CMS:2020ttz} & 13 TeV & $137$ fb$^{-1}$& $Hb,Zb$ & $0\ell$ \\
				& \cite{CMS:2024xbc} & 13 TeV & $138$ fb$^{-1}$ & Inclusive & $0\ell,2\ell$ SS \\
				& \cite{CMS:2015hyy} & 8 TeV & $19.7$ fb$^{-1}$& Inclusive & $1\ell,2\ell$ OS, $\geq3\ell$ \\
				& \cite{CMS:2018dcw} & 13 TeV & $35.9$ fb$^{-1}$ & $W^\pm t$ & $1\ell$ \\
				\hline
			\end{tabular}
		\end{adjustbox}
		\caption{Summary of CMS searches for the vector-like bottom partner. Here, $\ell=e,\mu$ and OS(SS) refers to the Opposite(Same) Sign case.}
		\label{tab:5}
	\end{table}
	
	\begin{table}[H]
		\centering
		\renewcommand{\arraystretch}{0.8}
		\begin{adjustbox}{max width=0.95\textwidth}
			\begin{tabular}{c|c|c|c|c|c}
				\hline\hline
				\textbf{VLQ} & \textbf{CMS} & $\sqrt{s}$ & \textbf{Luminosity} & \textbf{Decay} & \textbf{Lepton multiplicity} \\
				\hline\hline
				BR=100\% $X$ & \cite{CMS:2018dcw}&13 TeV & $35.9$ fb$^{-1}$ & $W^\pm t$ &$1\ell$\\
				& \cite{CMS:2018ubm} & 13 TeV & $35.9$ fb$^{-1}$ & $W^\pm t$ &$1\ell,2\ell$ SS\\
				\hline
				BR=100\% $Y$ & \cite{CMS:2017fpk} & 13 TeV & $2.3$ fb$^{-1}$ & $W^\pm(\ell\nu)b$ & $1\ell$ \\
				& \cite{CMS:2017ynm} & 13 TeV & $35.8$ fb$^{-1}$ & $W^\pm b$ & $1\ell$ \\
				\hline
				Triplet $Y$ & \cite{CMS:2024xbc} & 13 TeV & $138$ fb$^{-1}$ & Inclusive & $0\ell,2\ell$ SS \\
				\hline
			\end{tabular}
		\end{adjustbox}
		\caption{Summary of CMS searches for the exotic VLQs $X$ and $Y$. Here, $\ell=e,\mu$ and OS(SS) refers to the Opposite(Same) Sign case.}
		\label{tab:6}
	\end{table}
	
	\paragraph{Interpolation and Extrapolation Policy.}
	All experimental inputs are represented in their native published variables. \vlqb\ interpolates only inside the covered domain. No extrapolation is performed outside the available grid or curve range.
	
	\paragraph{Coupling-space Comparisons.}
	When an analysis publishes limits directly on $\kappa(m_Q)$, user inputs provided in terms of $s_{L/R}$ or $\Gamma_Q/m_Q$ are translated into $\kappa$ using Table~\ref{tab:kappa}. The comparison is then performed directly in the $(m_Q,\kappa)$ plane.
	
	\FloatBarrier
	\section{Software Architecture}
	\label{sec:architecture}
	
	The current \vlqb\ codebase follows a modern Python package layout and separates model definitions, physics utilities, experimental-data access, inference logic, and validation plotting. This structure is essential for reproducibility, because the validation figures and scan outputs are generated directly from the same internal data files and model classes used by the exclusion engine.
	The public source tree is intended to be distributed through GitHub at \githabrepo. In the final submission package, this repository will provide the tagged release, the machine-readable tables used by the code, and the example scripts employed to reproduce the validation plots and benchmark scans discussed in this work.
	
	\subsection{Design Goals}
	
	The main design goals are:
	\begin{itemize}[leftmargin=1.2cm]
		\item \emph{Data-driven evaluation}: experimental limits and theory tables are stored as machine-readable JSON inputs and loaded through the internal repository layer.
		\item \emph{Uniform model interface}: all supported VLQ representations expose common setters for mass, mixing, coupling, and width inputs.
		\item \emph{Native-limit interpretation}: cross-section limits, coupling limits, and width-scan limits are treated in their published variables.
		\item \emph{Transparent reporting}: the code returns the exclusion verdict, observed and expected ratios, dominant channel, process, experiment, luminosity, energy, and reference label.
		\item \emph{Direct validation plotting}: official-style validation plots can be generated directly from \vlqb, without embedding external numerical arrays in the user script.
	\end{itemize}
	
	\subsection{Package Map}

	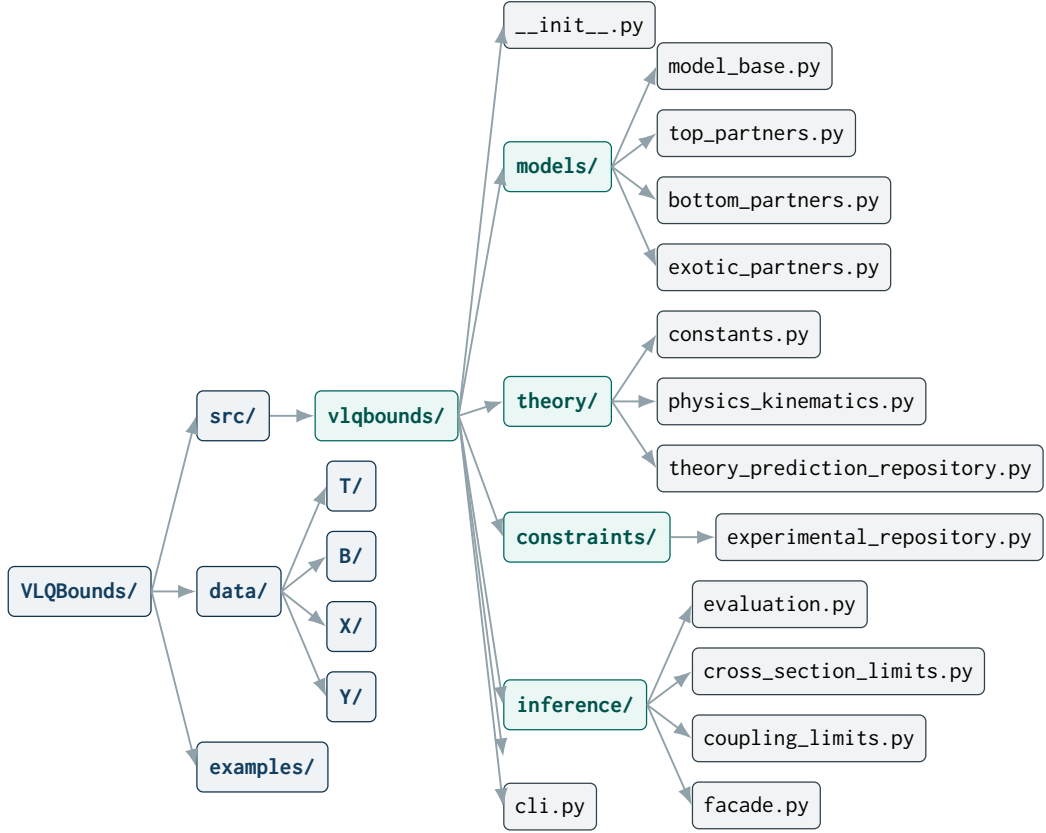
\begin{figure}[!htbp]
		\centering
		\resizebox{1.\linewidth}{!}{%
			\begin{forest}
				repository tree
				[\texttt{VLQBounds/}, repo folder
				[\texttt{src/}, repo folder
				[\texttt{vlqbounds/}, repo module
				[\texttt{\_\_init\_\_.py}, repo file]
				[\texttt{models/}, repo module
				[\texttt{model\_base.py}, repo file]
				[\texttt{top\_partners.py}, repo file]
				[\texttt{bottom\_partners.py}, repo file]
				[\texttt{exotic\_partners.py}, repo file]
				]
				[\texttt{theory/}, repo module
				[\texttt{constants.py}, repo file]
				[\texttt{physics\_kinematics.py}, repo file]
				[\texttt{theory\_prediction\_repository.py}, repo file]
				]
				[\texttt{constraints/}, repo module
				[\texttt{experimental\_repository.py}, repo file]
				]
				[\texttt{inference/}, repo module
				[\texttt{evaluation.py}, repo file]
				[\texttt{cross\_section\_limits.py}, repo file]
				[\texttt{coupling\_limits.py}, repo file]
				[\texttt{facade.py}, repo file]
				]
				[
				]
				[\texttt{cli.py}, repo file]
				]
				]
				[\texttt{data/}, repo folder
				[\texttt{T/}, repo folder]
				[\texttt{B/}, repo folder]
				[\texttt{X/}, repo folder]
				[\texttt{Y/}, repo folder]
				]
				[\texttt{examples/}, repo folder]
				]
				[\texttt{tests/}, repo folder]
				[\texttt{pyproject.toml}, repo file]
				[\texttt{README.md}, repo file]
				]
		\end{forest}}
		\caption{Layout of the \texttt{VLQBounds} repository.}
		\label{fig:package_map}
	\end{figure}
		Figure~\ref{fig:package_map} summarises the current package organisation. The installable package is located under \code{src/vlqbounds}. The \code{models} subpackage contains the VLQ classes, \code{theory} stores physics utilities and theory tables, \code{constraints} provides access to experimental repositories, \code{inference} contains the public \code{VLQBounds} facade, and \code{validation} contains the internal plotting tools for mass-width and mass-coupling validation figures.
	\subsection{Model Classes and Public Setters}
	
	The current model layer is based on the following classes:
	\begin{tcolorbox}[colback=VLQSoftBlue,colframe=VLQBlue!35!black,title=\textbf{Model class hierarchy}]
		\begin{lstlisting}[language=Python,numbers=none]
			VectorLikeModelBase
			TopPartnerSinglet
			TopPartnerDoublet
			TopPartnerPure
			BottomPartnerSinglet
			BottomPartnerDoublet
			BottomPartnerPure
			XPartnerDoublet
			YPartnerDoublet
			YPartnerTriplet
		\end{lstlisting}
	\end{tcolorbox}
	
	Each model class is called through the public \code{VLQBounds} facade. The supported parameter-setter methods are summarised in Table~\ref{tab:api_setters}. Each call requires a mass and one companion parameter: a mixing angle, an effective coupling, or a relative width.
	
	\begin{table}[H]
		\centering
		\renewcommand{\arraystretch}{1.12}
		\begin{tabular}{l|l|l}
			\hline\hline
			\textbf{Setter method} & \textbf{Mass keyword} & \textbf{Accepted companion keywords} \\
			\hline\hline
			\code{singletT\_params} & \code{mT} & \code{s\_l}, \code{k\_T}, \code{w\_m} \\
			\code{doubletT\_XT\_params} & \code{mT} & \code{s\_r}, \code{k\_T}, \code{w\_m} \\
			\code{doubletT\_TB\_params} & \code{mT} & \code{s\_u\_r}, \code{k\_T}, \code{w\_m} \\
			\hline
			\code{singletB\_params} & \code{mB} & \code{s\_l}, \code{k\_B}, \code{w\_m} \\
			\code{doubletB\_BY\_params} & \code{mB} & \code{s\_r}, \code{k\_B}, \code{w\_m} \\
			\code{doubletB\_TB\_params} & \code{mB} & \code{s\_d\_r}, \code{k\_B}, \code{w\_m} \\
			\hline
			\code{doubletX\_XT\_params} & \code{mX} & \code{s\_r}, \code{k\_X}, \code{w\_m} \\
			\hline
			\code{doubletY\_BY\_params} & \code{mY} & \code{s\_r}, \code{k\_Y}, \code{w\_m} \\
			\code{tripletY\_TBY\_params} & \code{mY} & \code{s\_d\_l}, \code{k\_Y}, \code{w\_m} \\
			\hline\hline
		\end{tabular}
		\caption{Public parameter-setter methods exposed by the \texttt{VLQBounds} facade.}
		\label{tab:api_setters}
	\end{table}
	
	\subsection{Execution Pipeline}
	
	\begin{tcolorbox}[colback=VLQSoftBlue,colframe=VLQBlue!35!black,title=\textbf{Deterministic decision flow}]
		\begin{enumerate}[leftmargin=1.0cm]
			\item \textbf{Model construction.} The user instantiates a model class from \code{vlqbounds.models} and passes it to \code{VLQBounds}.
			\item \textbf{Analysis preparation.} The method \code{prepare\_analysis()} loads cross-section inputs, coupling-limit tables, and the internal dictionaries required for the active VLQ species and representation.
			\item \textbf{Parameter assignment.} The user calls a typed setter such as \code{singletT\_params(mT=...,s\_l=...)}, \code{doubletB\_BY\_params(mB=...,w\_m=...)}, or \code{doubletT\_TB\_params(mT=...,k\_T=...)}.
			\item \textbf{Theory evaluation.} The active model computes masses, mixing angles, effective couplings, widths, BRs, and theory cross-sections.
			\item \textbf{Limit interpolation.} The inference layer interpolates the relevant observed and expected experimental limits in the native published variables.
			\item \textbf{Decision.} Channel-wise ratios are built using Eq.~\eqref{equ:obs_rat}. The most sensitive channel is selected and the exclusion verdict is stored.
			\item \textbf{Reporting.} Results are printed to the terminal, appended to the internal \code{pandas} DataFrame, and can be exported for scans or plotting.
		\end{enumerate}
	\end{tcolorbox}
	
	\subsection{Public API}
	
	A minimal single-point evaluation has the form:
	\Needspace{20\baselineskip}
	\begin{lstlisting}[language=Python,caption={Single-point evaluation with the current \texttt{VLQBounds} facade.}]
		from vlqbounds.models import TopPartnerSinglet
		from vlqbounds.inference import VLQBounds
		
		bounds = VLQBounds(TopPartnerSinglet())
		bounds.prepare_analysis()
		
		bounds.singletT_params(mT=1500.0, s_l=0.10)
		bounds.check_against_xs_and_coupling_limits()
		
		bounds.print_result()
		print(bounds.df.tail(1).T)
	\end{lstlisting}
	
	The same interface works for other species and representations:
	\Needspace{20\baselineskip}
	\begin{lstlisting}[language=Python,caption={Representative calls for different VLQ representations.}]
		from vlqbounds.models import (
		TopPartnerDoublet,
		BottomPartnerDoublet,
		XPartnerDoublet,
		YPartnerTriplet,
		)
		from vlqbounds.inference import VLQBounds
		
		# T doublet in the (T,B) representation
		bounds_t = VLQBounds(TopPartnerDoublet())
		bounds_t.prepare_analysis()
		bounds_t.doubletT_TB_params(mT=1200.0, k_T=0.35)
		bounds_t.check_against_xs_and_coupling_limits()
		
		# B doublet in the (B,Y) representation
		bounds_b = VLQBounds(BottomPartnerDoublet())
		bounds_b.prepare_analysis()
		bounds_b.doubletB_BY_params(mB=1000.0, w_m=0.05)
		bounds_b.check_against_xs_and_coupling_limits()
		
		# X doublet in the (X,T) representation
		bounds_x = VLQBounds(XPartnerDoublet())
		bounds_x.prepare_analysis()
		bounds_x.doubletX_XT_params(mX=1300.0, s_r=0.20)
		bounds_x.check_against_xs_and_coupling_limits()
		
		# Y triplet in the (T,B,Y) representation
		bounds_y = VLQBounds(YPartnerTriplet())
		bounds_y.prepare_analysis()
		bounds_y.tripletY_TBY_params(mY=1100.0, s_d_l=0.08)
		bounds_y.check_against_xs_and_coupling_limits()
	\end{lstlisting}
	
	\subsection{Scan Interface}
	
	The example scripts distributed with the package use a small support module to run scans and save tables and plots. For example, a scan over the $T$ singlet scenario is written as
	\Needspace{20\baselineskip}
	\begin{lstlisting}[language=Python,caption={Example scan script for the $T$ singlet scenario.}]
		import numpy as np
		
		from vlqbounds.models import TopPartnerSinglet
		from vlqbounds.inference import VLQBounds
		from example_support import run_parameter_scan, save_scan_outputs
		
		
		def configure(bounds, mass: float, mixing: float) -> None:
		bounds.singletT_params(mT=mass, s_l=mixing)
		
		
		def main() -> None:
		model = TopPartnerSinglet()
		bounds = VLQBounds(model)
		bounds.prepare_analysis()
		
		result_df = run_parameter_scan(
		bounds=bounds,
		parameter_callback=configure,
		masses=np.arange(600, 2001, 200),
		mixings=np.linspace(1e-5, 1.0, 10),
		)
		
		save_scan_outputs(
		result_df,
		output_stem="scan_t_singlet",
		x_label=r"$m_T\,\mathrm{[GeV]}$",
		title=r"$T\ \mathrm{Singlet}$",
		xlim=(500, 2000),
		ylim=(0.0, 0.5),
		)
		
		
		if __name__ == "__main__":
		main()
	\end{lstlisting}
	
	Similarly, the $B$ doublet in the $(B,Y)$ representation is scanned as
	\Needspace{20\baselineskip}
	\begin{lstlisting}[language=Python,caption={Example scan script for the $B$ doublet $(B,Y)$ scenario.}]
		import numpy as np
		
		from vlqbounds.models import BottomPartnerDoublet
		from vlqbounds.inference import VLQBounds
		from example_support import run_parameter_scan, save_scan_outputs
		
		
		def configure(bounds, mass: float, mixing: float) -> None:
		bounds.doubletB_BY_params(mB=mass, s_r=mixing)
		
		
		def main() -> None:
		model = BottomPartnerDoublet()
		bounds = VLQBounds(model)
		bounds.prepare_analysis()
		
		result_df = run_parameter_scan(
		bounds=bounds,
		parameter_callback=configure,
		masses=np.arange(600, 2001, 200),
		mixings=np.linspace(1e-5, 1.0, 10),
		)
		
		save_scan_outputs(
		result_df,
		output_stem="scan_b_doublet_by",
		x_label=r"$m_B\,\mathrm{[GeV]}$",
		title=r"$BY\ \mathrm{Doublet}$",
		)
		
		
		if __name__ == "__main__":
		main()
	\end{lstlisting}
	
	\subsection{Internal Validation Plotting}
	
	The validation plotting tools are part of \vlqb\ itself. The user script does not contain experimental or theory arrays; it only selects the internal limit key and output name.
	
	For the CMS mass-width scan of Ref.~\cite{CMS:2022yxp}, the validation figure is produced by
	\Needspace{20\baselineskip}
	\begin{lstlisting}[language=Python,caption={Internal CMS mass-width validation plot generated directly from \texttt{VLQBounds}.}]
		from vlqbounds.models import TopPartnerSinglet
		from vlqbounds.inference import VLQBounds
		
		bounds = VLQBounds(TopPartnerSinglet())
		bounds.prepare_analysis()
		
		surface, boundary = bounds.plot_width_scan_limit(
		limit_key="cms_2201_02227_t_singlet_tzbq",
		output_stem="validation_cms_2201_02227_t_singlet",
		)
	\end{lstlisting}
	
	For the ATLAS mass-coupling plot associated with Fig.~9 of Ref.~\cite{ATLAS:2022ozf}, the internal coupling-limit plotter is called as
	\Needspace{20\baselineskip}
	\begin{lstlisting}[language=Python,caption={Internal ATLAS mass-coupling validation plot generated directly from \texttt{VLQBounds}.}]
		from vlqbounds.models import TopPartnerSinglet
		from vlqbounds.inference import VLQBounds
		
		bounds = VLQBounds(TopPartnerSinglet())
		bounds.prepare_analysis()
		
		curves = bounds.plot_coupling_limit_figure(
		limit_key="atlas_2201_07045_fig9_t_singlet_ht",
		output_stem="validation_atlas_2201_07045_fig9_t_singlet_ht",
		use_tex=False,
		debug=True,
		)
	\end{lstlisting}
	
	The validation tools read the relevant internal JSON files, reconstruct the observed and expected curves, and save the corresponding numerical outputs and figures. In this way, validation plots and scan plots are produced within the same package workflow.
	
	\subsection{Command-line Usage}
	After editable installation,
	\Needspace{20\baselineskip}
	\begin{lstlisting}[language=bash,caption={Editable installation and example execution.}]
		python -m pip install -e .
		python examples/validation.py
		python examples/scan_t_singlet.py
	\end{lstlisting}
	
	\FloatBarrier
	\section{Validation}
	\label{sec:validation}
	
	\vlqb\ retrieves the theoretical predictions and experimental limits directly from the machine-readable files included in the repository. For a given point, specified by the VLQ mass and either a coupling-like input or a relative width $\Gamma_Q/m_Q$, the framework evaluates the corresponding theory cross-section, interpolates the experimental upper limit, and constructs the sensitivity ratios of Eq.~\eqref{equ:obs_rat}.
	
\begin{figure}[!htbp]
		\centering
		\safeincludegraphics[width=0.7\linewidth]{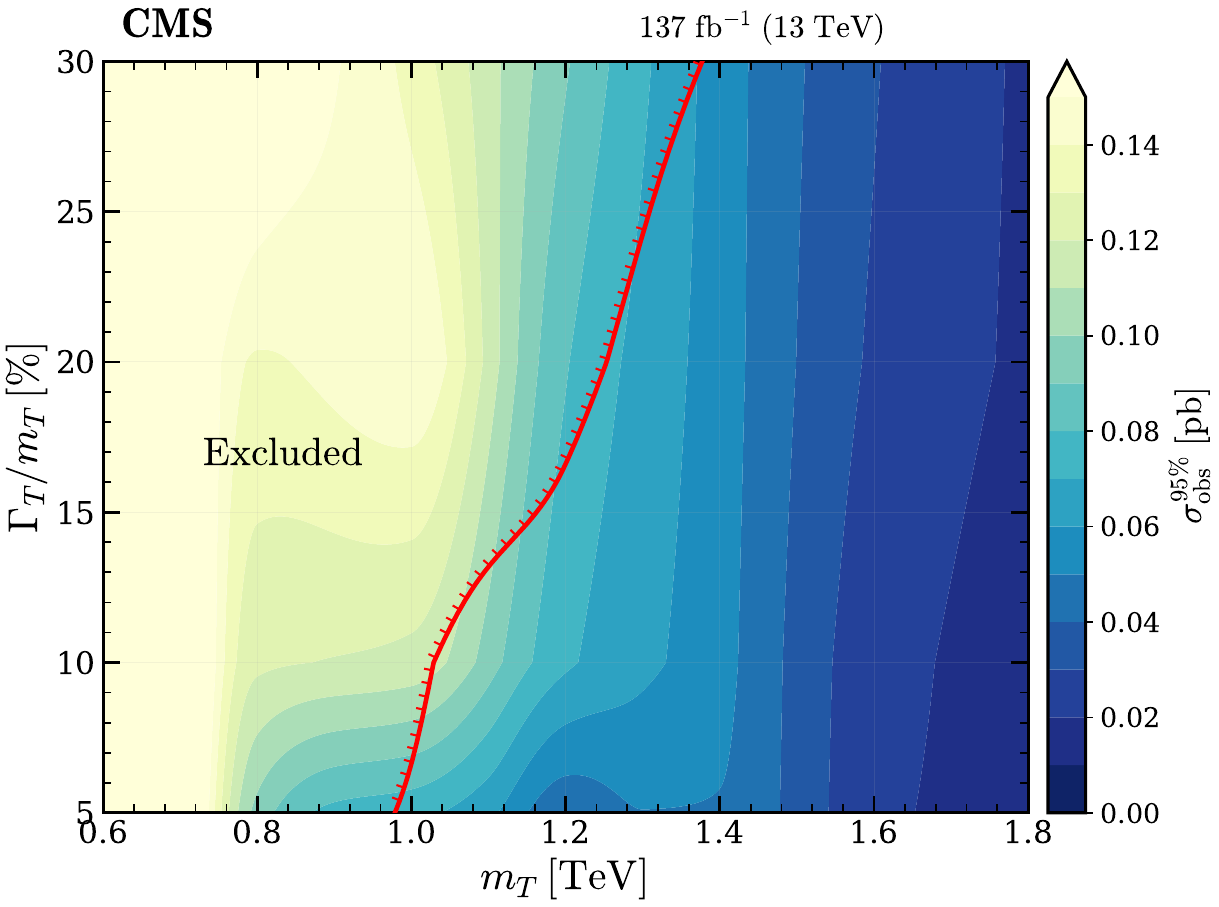}
		\caption{Observed 95\% CL upper limit on $\sigma(pp\to Tbq\to tZbq)$ in the $(m_T,\Gamma_T/m_T)$ plane.}
		\label{fig:cms_interp}
	\end{figure}

	The CMS validation shown in Fig.~\ref{fig:cms_interp} uses the published width slices $\Gamma_T/m_T=5\%,10\%,20\%,$ and $30\%$ for the singlet $T$ search in the $tZbq$ final state~\cite{CMS:2022yxp}. The internal validation routine performs a structured interpolation: shape-preserving interpolation in mass within each published width slice, followed by interpolation in width. The red curve is defined by $r_{\rm obs}=1$, with the excluded region lying on the low-mass side of the contour.
	
\begin{figure}[!htbp]
		\centering
		\safeincludegraphics[width=0.7\linewidth]{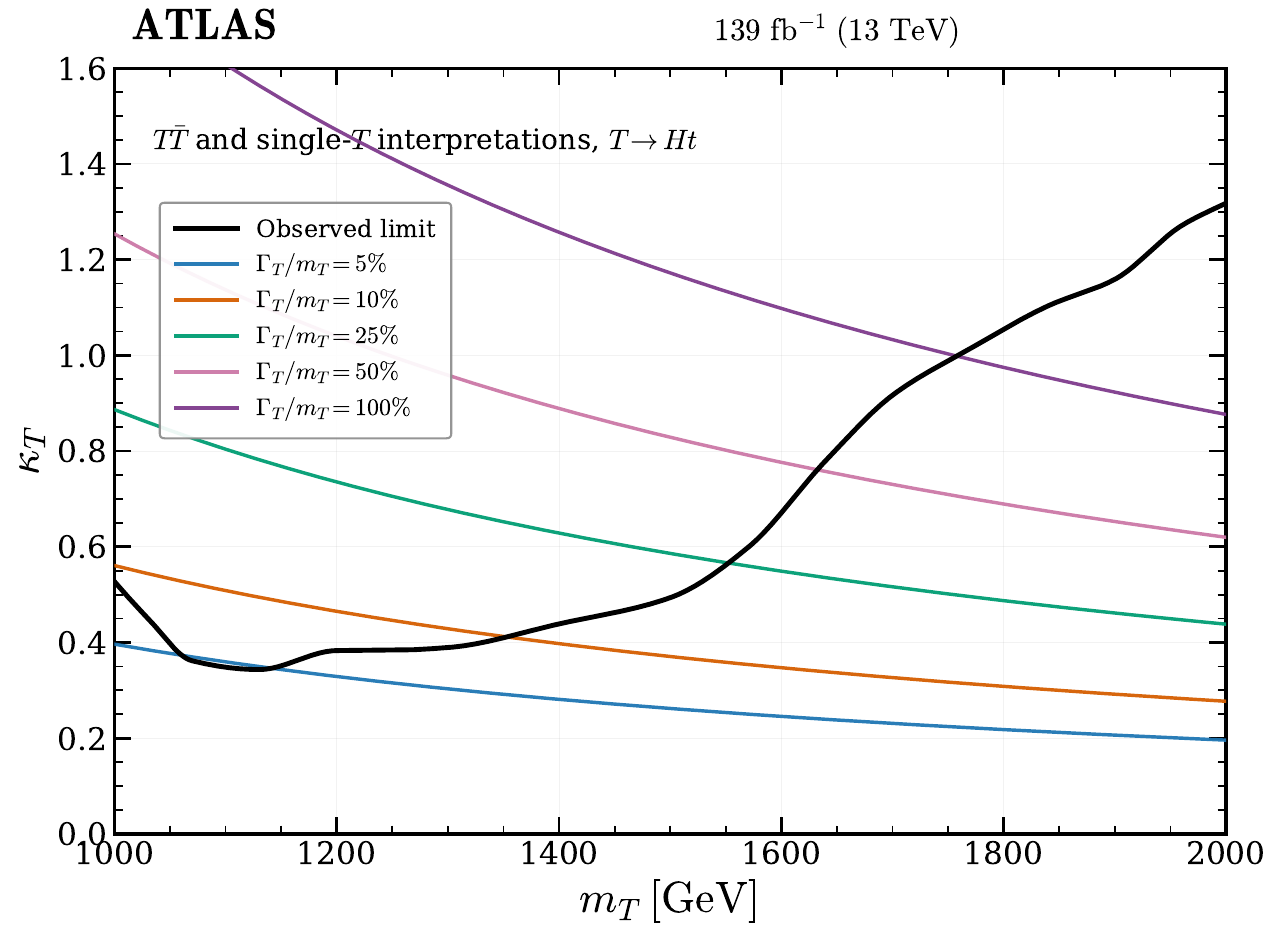}
		\caption{Observed and expected upper limits on the effective coupling $\kappa_T$ as a function of the vector-like top mass.}
		\label{fig:atlas_interp}
	\end{figure}

	Figure~\ref{fig:atlas_interp} shows the corresponding coupling-limit validation for the ATLAS analysis of Ref.~\cite{ATLAS:2022ozf}. In this case, \vlqb\ reconstructs the observed and expected limits in the $(m_T,\kappa_T)$ plane from the internal digitised data files. The same interface also overlays reference curves corresponding to fixed values of $\Gamma_T/m_T$, computed from the active \code{TopPartnerSinglet} model.
	
	These validation examples test both the numerical interpolation and the internal mapping between user-level input parameters and the native variables used by the experimental searches. They also provide direct regression tests for the plotting layer, since the figures are generated using the same data repository and model interface used by the exclusion engine.
	
	\FloatBarrier
	\section{Representative Results}
	\label{sec:results}
	
	\begin{figure}[!htbp]
		\centering
		\begin{minipage}{0.48\linewidth}
			\centering
			\safeincludegraphics[width=\linewidth]{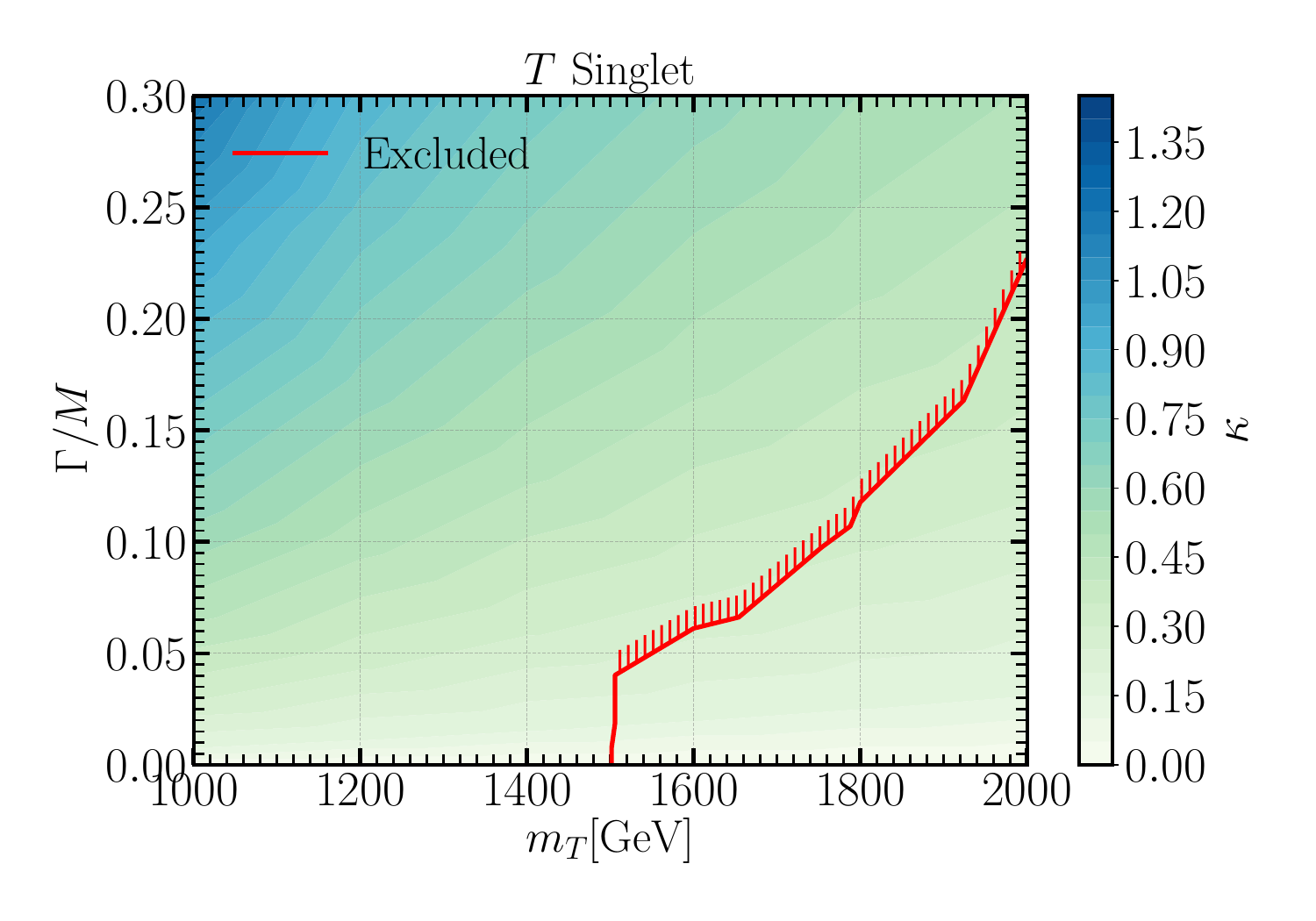}
		\end{minipage}\hfill
		\begin{minipage}{0.48\linewidth}
			\centering
			\safeincludegraphics[width=\linewidth]{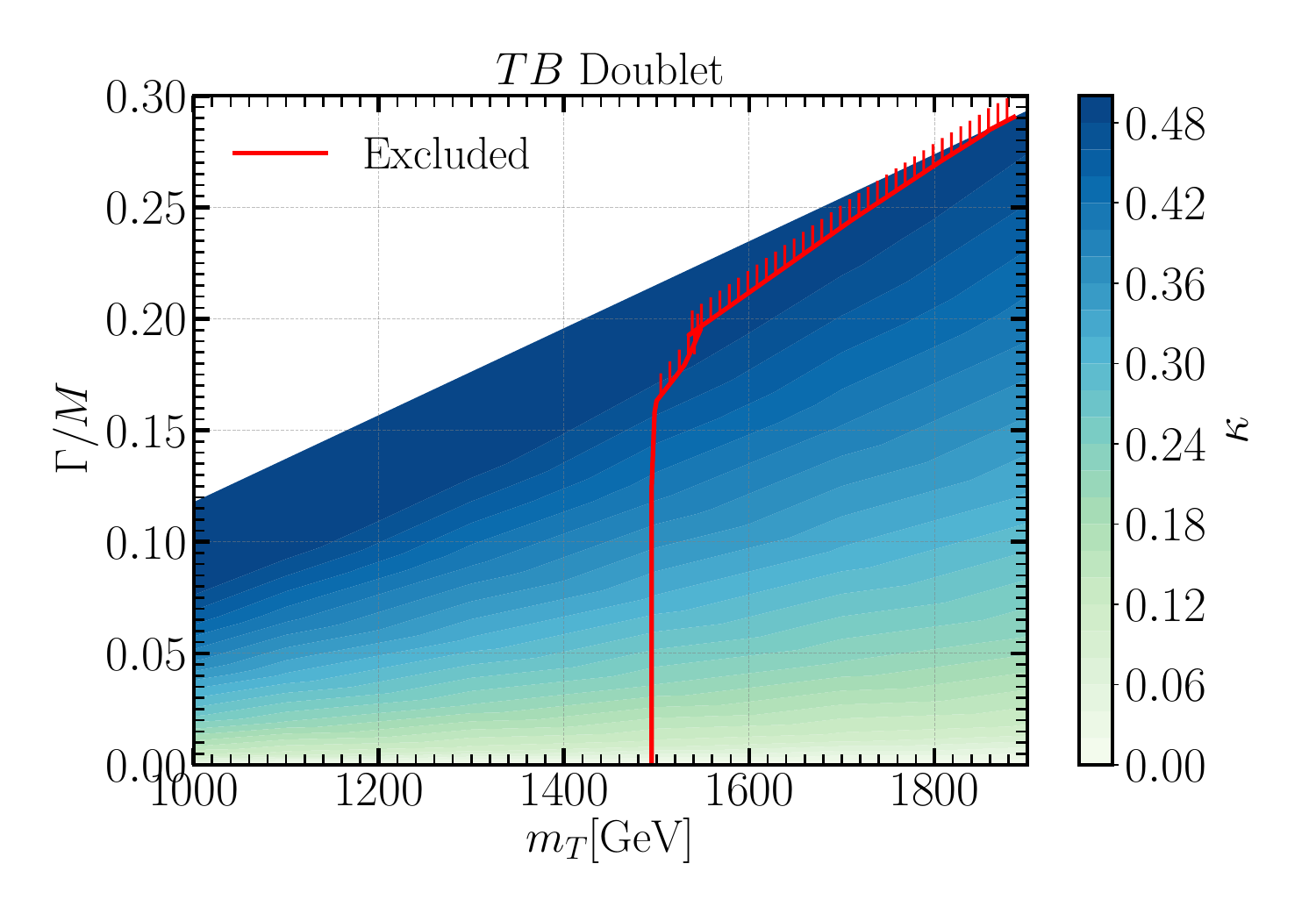}
		\end{minipage}
		
		\vspace{2mm}
		
		\begin{minipage}{0.48\linewidth}
			\centering
			\safeincludegraphics[width=\linewidth]{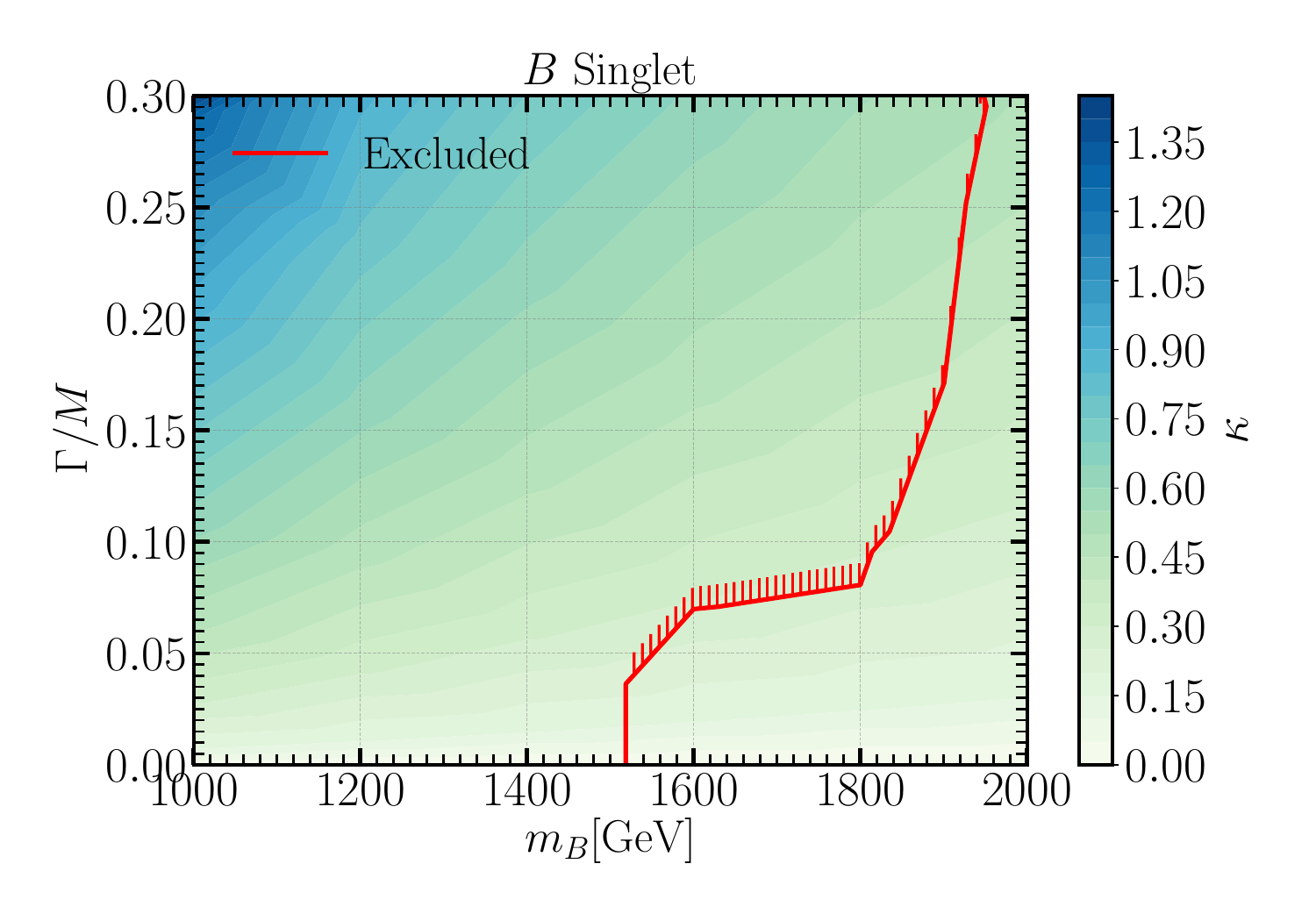}
		\end{minipage}\hfill
		\begin{minipage}{0.48\linewidth}
			\centering
			\safeincludegraphics[width=\linewidth]{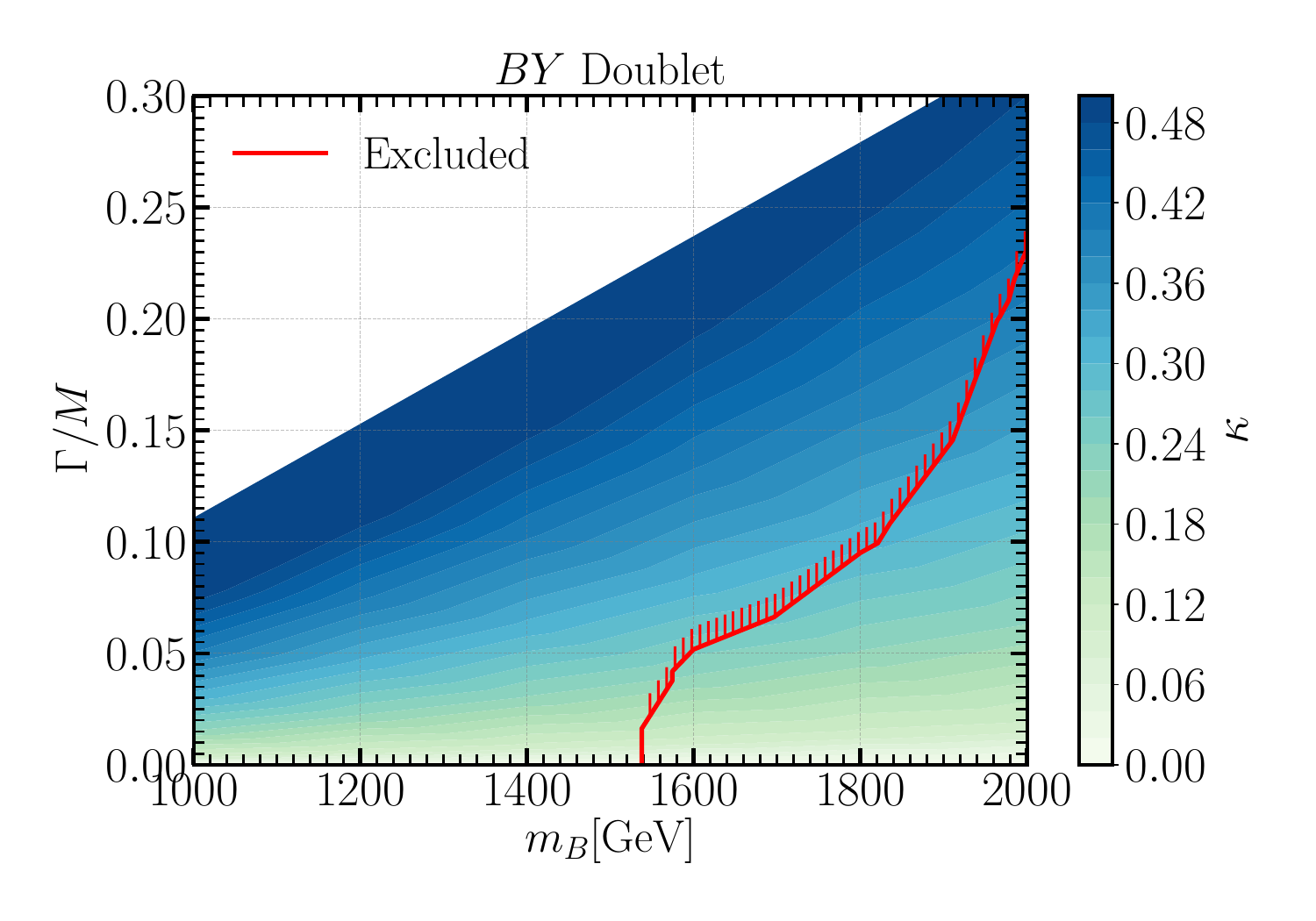}
		\end{minipage}
		
		\vspace{2mm}
		
		\begin{minipage}{0.48\linewidth}
			\centering
			\safeincludegraphics[width=\linewidth]{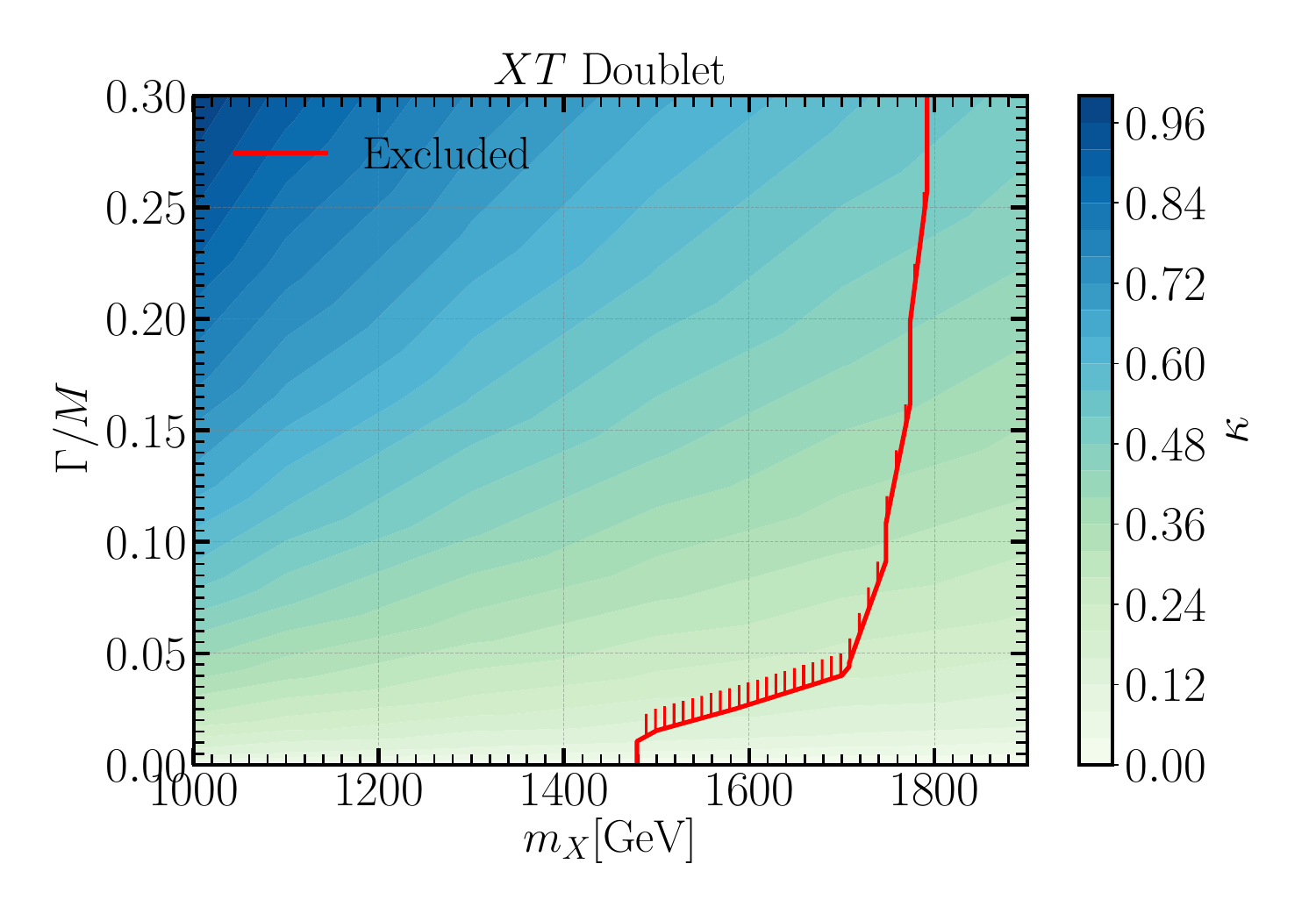}
		\end{minipage}\hfill
		\begin{minipage}{0.48\linewidth}
			\centering
			\safeincludegraphics[width=\linewidth]{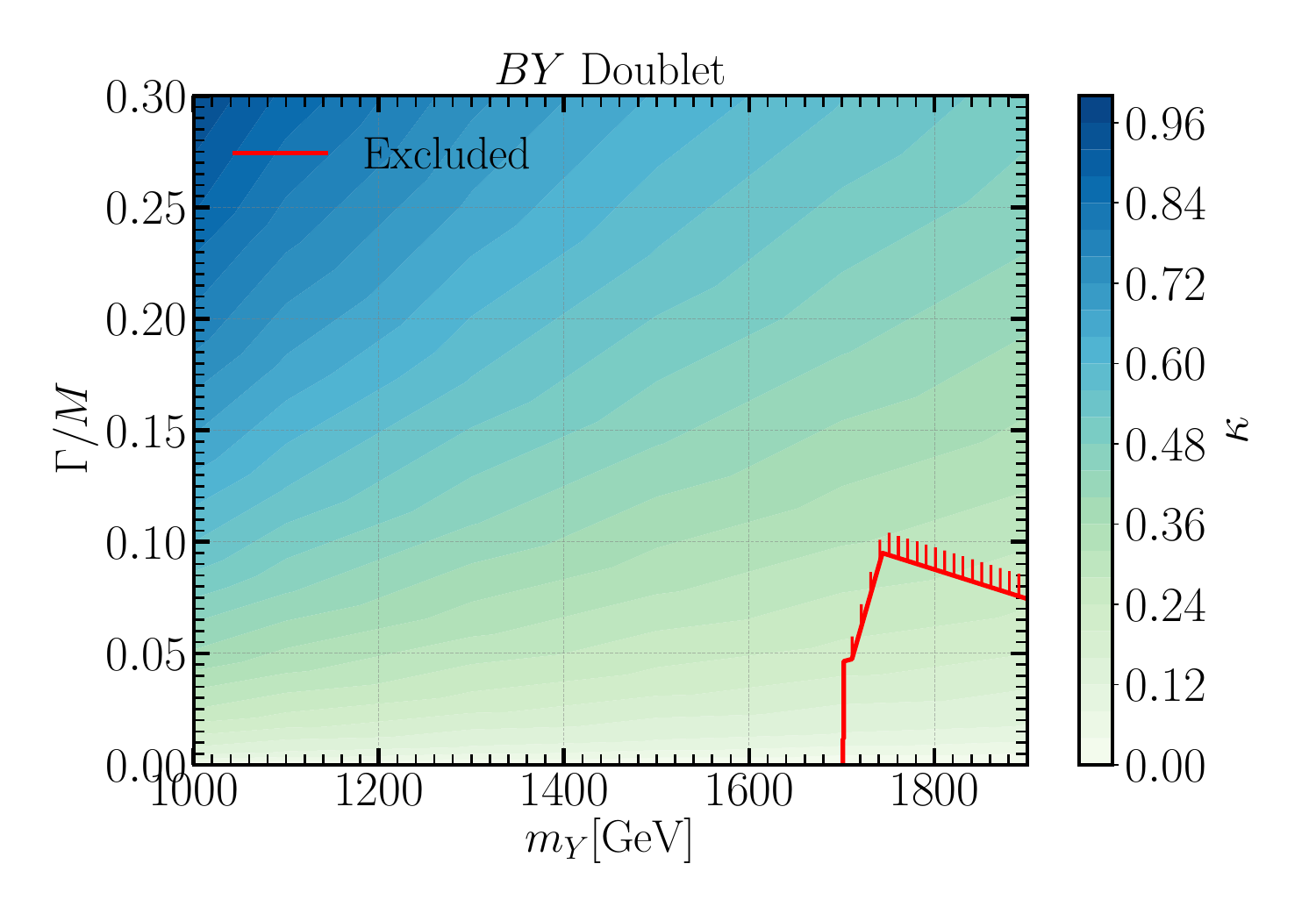}
		\end{minipage}
		\caption{Observed 95\% CL exclusion contours for six representative VLQ benchmark scenarios produced with \texttt{VLQBounds}.}
		\label{fig:benchmarks}
	\end{figure}
	
	We illustrate the workflow using six representative benchmark scenarios. For each scenario, the scan is performed in the $(m_Q,\Gamma_Q/m_Q)$ plane. The colour map reports the corresponding effective coupling $\kappa$, while the red contour denotes the observed 95\% CL exclusion boundary after selecting the most sensitive analysis at each parameter point.
	
	A typical structured output produced by \vlqb\ is shown in Listing~\ref{list:min_key}. The output records the model definition, the input parameters, derived quantities, the exclusion verdict, the most sensitive channel, and the metadata needed for reproducibility.
	
	\Needspace{20\baselineskip}
	\begin{lstlisting}[caption={Illustrative structured output produced by \texttt{VLQBounds}.},label={list:min_key}]
		{
			"scenario": {
				"model": "TopPartnerSinglet",
				"input": {
					"mass_GeV": 1200.0,
					"s_left": 0.4003
				},
				"derived": {
					"width_over_mass": 0.1397,
					"effective_kappa": 0.5661,
					"br_Wb": 0.501,
					"br_Zt": 0.249,
					"br_Ht": 0.250
				}
			},
			"result": {
				"decision": "Excluded",
				"result_flag": 0,
				"observed_ratio": 5.00,
				"expected_ratio": 4.42,
				"dominant_channel": 45
			},
			"driving_analysis": {
				"description": "See manifest metadata for the full experimental record",
				"note": "Most sensitive channel selected through expected sensitivity."
			},
			"metadata": {
				"vlqbounds_version": "0.1",
				"generated_by": "VLQBounds runtime",
				"data_format": "JSON"
			}
		}
	\end{lstlisting}
	
	Figure~\ref{fig:benchmarks} shows a common qualitative pattern. At small relative widths, exclusions are dominated by QCD pair-production searches, producing near-vertical mass thresholds. As the width increases, and hence as the effective coupling grows, single-production channels become increasingly important and extend the sensitivity to larger masses. Differences between singlet and doublet scenarios arise from the representation-dependent mapping between the mixing angle and $\kappa$, as well as from changes in the BRs.
	
	\section*{Program summary (\textit{Comput.\ Phys.\ Commun.})}
	
	\begin{tcolorbox}[colback=VLQSoftBlue,colframe=VLQBlue!35!black]
		\textbf{Program title:} \vlqb\\
		\textbf{Licensing provisions:} open source.\\
		\textbf{Repository:} \githabrepo\\
		\textbf{Programming language:} Python 3.10+; plotting based on \texttt{matplotlib}.\\
		\textbf{Nature of problem:} determine whether a user-specified VLQ hypothesis is excluded at 95\% CL by public ATLAS and CMS searches.\\
		\textbf{Solution method:} JSON-based loading of experimental inputs; conversion between mixing, effective-coupling, and relative-width parameterisations; calculation of partial widths and branching ratios; interpolation of cross-section and coupling limits; identification of the most sensitive channel; and reporting of results in terminal and tabular form.\\
		\textbf{Restrictions:} parameter points outside the published experimental coverage are flagged and are not extrapolated.
	\end{tcolorbox}
	
	\FloatBarrier
	\section{Conclusions}
	\label{sec:conclusion}
	
	We have presented \vlqb, a Python framework for confronting VLQ scenarios with public ATLAS and CMS searches. The code accepts minimal user input -- the VLQ species and representation, the heavy mass, and one of $s_{L/R}$, $\kappa$, or $\Gamma_Q/m_Q$ -- and maps this information onto the native parameterisation of the relevant experimental limits. It then computes the corresponding production and decay information, performs interpolation within the published domains, identifies the most sensitive analysis, and returns a 95\% CL exclusion verdict together with diagnostic observables.
	
	The implementation is deliberately conservative. Experimental information is used only within the published kinematic coverage, with no extrapolation beyond available grids. At the same time, the code is designed for reproducibility and extensibility: all results are traceable to machine-readable internal data files, and validation plots are generated directly through the same package interface used for physical scans.
	
	Although the present implementation already covers a broad set of public VLQ limits, the structure of \vlqb\ naturally accommodates future extensions. A particularly important next step is the extension from the present SM$+$VLQ implementation to a 2HDM$+$VLQ framework, where exotic decay channels such as $T\to Ht$, $T\to At$, $T\to H^+b$, $B\to Hb$, $B\to Ab$, $B\to H^-t$, together with the corresponding modes involving the exotic $X$ and $Y$ states, become central ingredients of the phenomenology~\cite{Arhrib:2024dou,Arhrib:2024tzm,Arhrib:2024mbq,Benbrik:2022kpo,Abouabid:2023mbu,Benbrik:2024bxt,Benbrik:2024hsf,Arhrib:2024nbj,Benbrik:2023xlo,Benbrik:2025nfw}. Further extensions include additional single-production channels, Run-3 and High-Luminosity LHC projections, and future lepton-collider setups. 
	
	\section*{Acknowledgments}
	\sloppy
	M. Boukidi acknowledges the support of Narodowe Centrum Nauki under OPUS grant no.\ 2023/49/B/ST2/03862. SM is supported in part through the NExT Institute and STFC CG ST/X000583/1.
	
	\bibliographystyle{elsarticle-num}
	\bibliography{main,vlqbounds_additions}
	
\end{document}